\documentclass{aa}  

\usepackage{graphicx}	
\usepackage{amsmath}	
\usepackage{amssymb}	
\usepackage[flushleft]{threeparttable}
\usepackage[stable]{footmisc}

\def\mbh{\hbox{$M_{\rm BH}$}}

\def\spin{\hbox{$a^\ast$}}

\newcommand{\mdot}{\ensuremath{\dot{m}/\dot{m}_{\rm Edd}}}
\newcommand{\rg}{\ensuremath{r_{\rm g}}}
\newcommand{\ltransf}{\ensuremath{L_{\rm transf}/L_{\rm disc}}}
\newcommand{\fcol}{\ensuremath{f_{\rm col}}}

\def\kynsed{{\tt KYNSED}}

\def\swift{\textit{Swift}}
\def\hst{\textit{HST}}
\usepackage[breaklinks, colorlinks, citecolor=blue, urlcolor = blue]{hyperref}

\makeatletter
\renewcommand*\aa@pageof{, page \thepage{} of \pageref*{LastPage}}
\makeatother

\begin{document} 


\title{Broadband X-ray/UV/optical time-resolved spectroscopy of NGC\,5548: The origin of the UV/optical variability in active galactic nuclei}    
\titlerunning{Broad band, time-resolved spectroscopy of NGC\,5548 }
    
\author{E. Kammoun \inst{\ref{inst1},\ref{inst2},\ref{inst3}}
\and
I. E. Papadakis \inst{\ref{inst4},\ref{inst5}}
\and
M. Dov{\v c}iak \inst{\ref{inst6}}
\and
C. Panagiotou \inst{\ref{inst7}}
}
\institute{
Dipartimento di Matematica e Fisica, Universit\`{a} Roma Tre, via della Vasca Navale 84, I-00146 Rome, Italy \email{\href{mailto:ekammoun.astro@gmail.com}{ekammoun.astro@gmail.com}}\label{inst1}
\and
IRAP, Universit\'e de Toulouse, CNRS, UPS, CNES 9, Avenue du Colonel Roche, BP 44346, F-31028, Toulouse Cedex 4, France \label{inst2}
\and
INAF -- Osservatorio Astrofisico di Arcetri, Largo Enrico Fermi 5, I-50125 Firenze, Italy\label{inst3} 
\and
Department of Physics and Institute of Theoretical and Computational Physics, University of Crete, 71003 Heraklion, Greece \label{inst4} 
\and 
Institute of Astrophysics, FORTH, GR-71110 Heraklion, Greece\label{inst5} 
\and 
Astronomical Institute of the Czech Academy of Sciences, Bo{\v c}n{\'i} II 1401, CZ-14100 Prague, Czech Republic  \label{inst6}
\and
MIT Kavli Institute for Astrophysics and Space 2 Research, Massachusetts Institute of Technology, Cambridge, MA 02139, USA \label{inst7} 
}

\date{Received ; accepted }

 
  \abstract{Recently, nearby bright galaxies have been the subject of long monitoring surveys in the X-rays, UV, and optical. All of these campaigns revealed a strong correlation between the various UV and optical bands, with time lags that increase with wavelength. In a series of papers, we have demonstrated that a scenario in which an X-ray source located on the rotation axis of the central black hole illuminating the accretion disc is a viable explanation for the observed correlations. However, some of the monitored sources showed low or moderate correlation between the X-rays and the UV, which could challenge this scenario.}
  {In this paper, we analyse the time-averaged and the variable broadband X-ray/UV/optical spectral energy distributions (SEDs) of NGC\,5548, one of the most intensely monitored Seyfert\,1 galaxies, using \swift, \hst, and ground-based telescopes. The aim of this paper is to test whether the broadband spectral behaviour of the source could be explained with the X-ray illumination hypothesis, despite the apparently moderate correlation between the X-rays and longer wavelength.}
  {We modelled the broadband time-averaged SED, from the Space Telescope and Optical Reverberation Mapping (STORM) monitoring campaign of the source performed using the {\tt KYNSED} model, which assumes X-ray illumination of the disc. We also modelled 15 time-resolved SEDs extracted from the same campaign to check whether this model can account for the spectral variability seen in the various wavebands during the monitoring. We assumed in our modelling that the X-ray corona is powered via the accretion process.}
  {Our results show that the proposed scenario could describe the time-averaged and the time-resolved SEDs of NGC\,5548 perfectly well. In this scenario, the height of the corona, the X-ray photon index, and the power that is transferred to the corona all vary. This would explain the variability behaviour at the various wavelengths. The best-fit model is obtained for a non-spinning black hole accreting at a constant rate of 5\% of its Eddington limit. Since each of the variable parameters will affect the observed flux in a particular way, the combined variability of all of these parameters will then account for the moderate correlation between the X-rays and UV/optical.}
  {We demonstrate in this paper that X-ray illumination of the accretion disc can actually explain the observed properties of NGC\,5548. In fact, this model not only fits the broadband spectra of the source well, but it also explains the time-lag behaviour as a function of wavelength as well as the power spectral distribution, providing a complete description of the behaviour of this source.} 

   \keywords{Accretion, accretion discs -- Galaxies: active -- Galaxies: Seyfert }

   \maketitle
%

\section{Introduction}
\label{sec:intro}

Active galactic nuclei (AGN) are the most luminous persistent sources of electromagnetic radiation in the universe. It is commonly accepted that these objects are powered by accretion onto a supermassive black hole (SMBH) in the form of an optically thick and geometrically thick accretion disc \citep[][]{Shakura73, Novikov73}. AGN are also known to be strong X-ray emitters. However, the nature and the geometry of the X-ray source (known as the X-ray corona) remains unknown. Owing to the difficulties in spatially resolving the accretion disc and the X-ray corona using current facilities, various techniques have been proposed to study the innermost regions of the accretion flow. For example, gravitational microlensing has been used to study the structure of the accretion disc and the X-ray corona \citep[e.g.][]{Pooley07, Chartas09, Mosquera13, Blackburne14, Chartas16}. Alternatively, reverberation mapping techniques have been widely used to study the different components of AGN. This technique uses the time lags between light curves observed at different wavelengths to infer the geometry and kinematics of the regions emitting at those wavelengths \citep[see][for a recent review]{Cackett21review}. 

Originally, this technique is used to measure the lags between the light curve in the H$\beta$ emission line and the UV continuum, aiming at studying the size scale and the dynamics of the broad line region \citep[BLR; e.g.][]{Blandford82, Peterson93, Peterson2004}. A variant of this technique was later used to study time delays between the variations in various X--ray wavebands. The objective was to constrain the geometry of the X--ray source with respect to the accretion disc, via the study of time-lags spectra between bands dominated by the X--ray continuum emission and those bands where the X--ray reflection features are dominant \citep[e.g.][]{Fabian2009, DeMarco2013, Kara2016, Emmanoulopoulos2011, Epitropakis16, Caballero18, Caballero20}.

More recently, a huge observational effort has been invested in long monitoring campaigns of many Seyfert galaxies using ground-based and space-based observatories in order to investigate the correlated (or not) variations of AGN in the X--rays and the UV/optical bands \citep[see e.g.][]{Derosa15, Fausnaugh16, Mchardy18, Cackett18, Edelson19, Cackett20, Santisteban20, Pahari20, Kara21, Cackett2023, Miller2023, Kara2023, Kumari23, Kumari24}. The main objective is to test theoretical models for the interplay between the X--ray and the fast UV/optical variations in AGN. For example, in the case when the central X-ray corona illuminates the accretion disc,  most of the X-rays are absorbed by the disc, which adds to its heating. Since the X-rays are highly variable in AGN, this mechanism drives variations in the thermally reprocessed continuum emission of the disc. These variations are seen with some delay which depends on the location of the X-rays above the disc as well as the BH mass, the disc accretion rate, etc. 

A standard \cite{Shakura73} accretion disc gives rise to wavelength-dependent lags following $\tau \propto \lambda^{4/3}$ \citep[see][and references therein]{Cackett07}, and the cross-correlation results from the  monitoring campaigns confirmed this prediction. However, it was originally thought that the amplitude of this relation  underestimated the observed lags by a factor of $\sim 3-4$. Recently, we published the results from a detailed study of X--ray thermal reverberation in a series of papers. \citet[][K21 hereafter]{Kammoun21a} studied in, detail, the response of a \cite{Novikov73} disc to X-ray illumination by a point-like source, located above the central BH, taking all relativistic effects into account. \cite{Kammoun19lags,Kammoun21b} showed that, when treated properly, the model time lags in the case of X-ray-illuminated discs are in perfect agreement with the observations, contrary to the previous claims. In addition, the K21 model can also explain the observed power spectra in the optical/UV bands of NGC\,5548 \citep{Panagiotou20, Panagiotou22}.

Recently, \citet[][K23 hereafter]{Kammoun23} published a new code that can be used to fit the observed time-lags when X--rays illuminate the disc, both in the case when  the X--ray corona is powered by the accretion process, and when it is powered externally, by an unknown source not directly associated with the accretion power. In the former case, the model assumes that the accretion power dissipated in the disc below a transition radius, $r_{\rm tr}=R_{\rm tr}/$\rg\footnote{\rg$=G$\mbh$/c^2$ is the gravitational radius of a BH with a mass of \mbh.}, is transferred to the X-ray source (by a yet unspecified physical mechanism). In this way, a direct link between the accretion disc and the X--ray source is set. They also considered various colour-correction factors, $f_{col}$, for the disc emission, and they presented results from fits to the time-lags spectra of four sources. In fact, electron scattering is known to play an important role and can lead to deviations from blackbody emission. In general, transfer of energy between electrons and photons that are Compton scattered enforces a Wien tail at the high energy end of the spectrum. The resulting spectra can be approximately modelled by modifying the blackbody spectrum using \fcol\ as a multiplicative factor by which spectral features are shifted to higher energies, while the normalisation of the spectrum is multiplied by \fcol$^{-4}$ to keep the frequency integrated flux fixed. In this case the modified blackbody spectrum can be expressed as follows,
\begin{equation}
    I_\nu = \frac{2h}{c^2\fcol^4}\frac{\nu^3}{\exp(\frac{h\nu}{\fcol kT})-1},
\end{equation}

\noindent where $I_\nu$ is the specific intensity, $k$ is Boltzmann’s constant, $h$ is Planck’s constant, and $T$ is the disc temperature. 

Time-lag spectra (i.e. distribution of time lags as a function of wavelength) are just one of the many tests one can perform in order to test a theoretical model for the X--ray/UV/optical emission in AGN. The available data from the long, multi-wavelength light curves can be used to test models in various ways. We aim to explore two of them, using data from the original Space Telescope and Optical Reverberation Mapping (STORM) multi-wavelength monitoring campaign in 2014 \citep{Derosa15, Fausnaugh16}. NGC\,5548 is, perhaps, the best target for these tests at the moment due to the quality of the available light curves (which are long and very well sampled) but also because the host galaxy emission and the `small-blue-bump' emission (i.e. the blended Fe{\tt II} with the Balmer continuum components) are well established in this source. 

Firstly, we use the spectral model \kynsed\, to fit the average SED of the source.  \kynsed\, was developed by \cite{Dovciak22} (D22 hereafter) to calculate the broadband optical/UV/X-ray spectral energy distribution (SED) of an AGN, assuming a \cite{Novikov73} accretion disc which is illuminated by a point-like X-ray corona. The model can provide the theoretical, broad band SED both in the case when the X--ray corona is powered by an unknown source as well as in the case when it is powered by the accretion process itself, similar to the case of the time lag model described above. 
In the latter case, a direct link between the accretion disc and the X--ray source is set. D22 already showed that this model can fit well the mean SED of NGC\,5548. Our aim is to investigate whether the model can fit well the mean SED assuming the best-fit results of \cite{Kammoun23}. In this way, we will be able to demonstrate that the model can fit well both the average SED and the average time-lags spectra in this source.

Secondly, we construct numerous SEDs, sampling X--ray, UV and optical data throughout the campaign and investigate whether X--ray reverberation can indeed explain the complex, X--ray/UV/optical spectral variability of the source. The second objective aims at directly addressing a puzzling result from the cross-correlation analysis of the X--ray and the UV/optical light curves. The STORM light curves of NGC\,5548 implied a rather low cross-correlation amplitude between the X--ray and the UV/optical light curves. Even by visual inspection, the shape of the X-ray light curve did not appear capable of driving the UV/optical variability \citep{Gardner17, Starkey17}. \cite{Panagiotou22} already showed that an apparent low cross-correlation between the X--ray and the UV/optical light curves can be explained in a physical way within the X--ray reverberation model. This work shows that, indeed, the multiwavelength light curves of NGC\,5548 are fully consistent with the assumption of X--ray thermal reverberation being the only driver of the observed variations in the UV/optical bands.  

The paper is organised as follows. In Sect.\,\ref{sec:average spectrum}, we present the fitting of the time-averaged SED of NGC\,5548. In Sect.\,\ref{sec:variableSED} we present the analysis of the time-variable SED of NGC\,5548. Finally, we summarise our results and discuss the main implications of our work in Sect.\,\ref{sec:discussion}.

\section{Fitting the average SED}
\label{sec:average spectrum}

We use the same data set presented in D22 as described in their Section\,4.1 to construct the mean SED. We also adopt the same model used in D22 to it the data. The overall model, written in {\tt XSPEC} \citep{Arnaud1996} parlance, is  as follows,
\begin{eqnarray}
\rm Model &=& {\rm \tt extinction_{Czerny} \times \kynsed_{UV/opt.} } \nonumber \\
& + & {\rm \tt TBabs \times zTBabs \times zxipcf  \times \kynsed_{\rm X-rays}}.
\label{xspecmdel}
\end{eqnarray}

\noindent We use \kynsed\, twice; one for the optical/UV and the other for the X-ray spectra. Obviously, all the relevant \kynsed\, parameters were tied in the UV/optical and X-rays range. The {\tt TBabs} component \citep{wilms00} accounts for the effects of the Galactic absorption in the X--ray band (the UV/optical data points are already corrected for Galactic extinction). We fix its column density to $1.55\times 10^{20}~\rm cm^{-2}$ \citep{HI4PI}.

We also consider the possibility of intrinsic absorption in the host galaxy. We assume the extinction curve of \cite{Czerny04} for the absorption in the optical/UV. The {\tt extinction$_{\rm Czerny}$} component is determined by $E(B-V)_{\rm host}$ which is left as a free parameter of the fit. The {\tt zTBabs} component accounts for the absorption in the X--ray band of the same absorber, and its column density was kept linked to $E(B-V)_{\rm host}$  through the relation $N_{\rm H, host} =  8.3 \times 10^{21}~E(B-V)_{\rm host} ~ {\rm cm^{-2}}$ \citep{Liszt2021}. Finally, we use the spectral component {\tt zxipcf} \citep{Reeves2008}, to account for possible warm absorption. This component accounts for absorption from partially ionised absorbing material, which covers some fraction of the source, while the remaining of the spectrum is seen directly. First, we let the covering fraction of this component free, but it turned out to be consistent with one. So we fix it at this value.

\begin{figure}
    \centering
    \includegraphics[width = 1\linewidth]{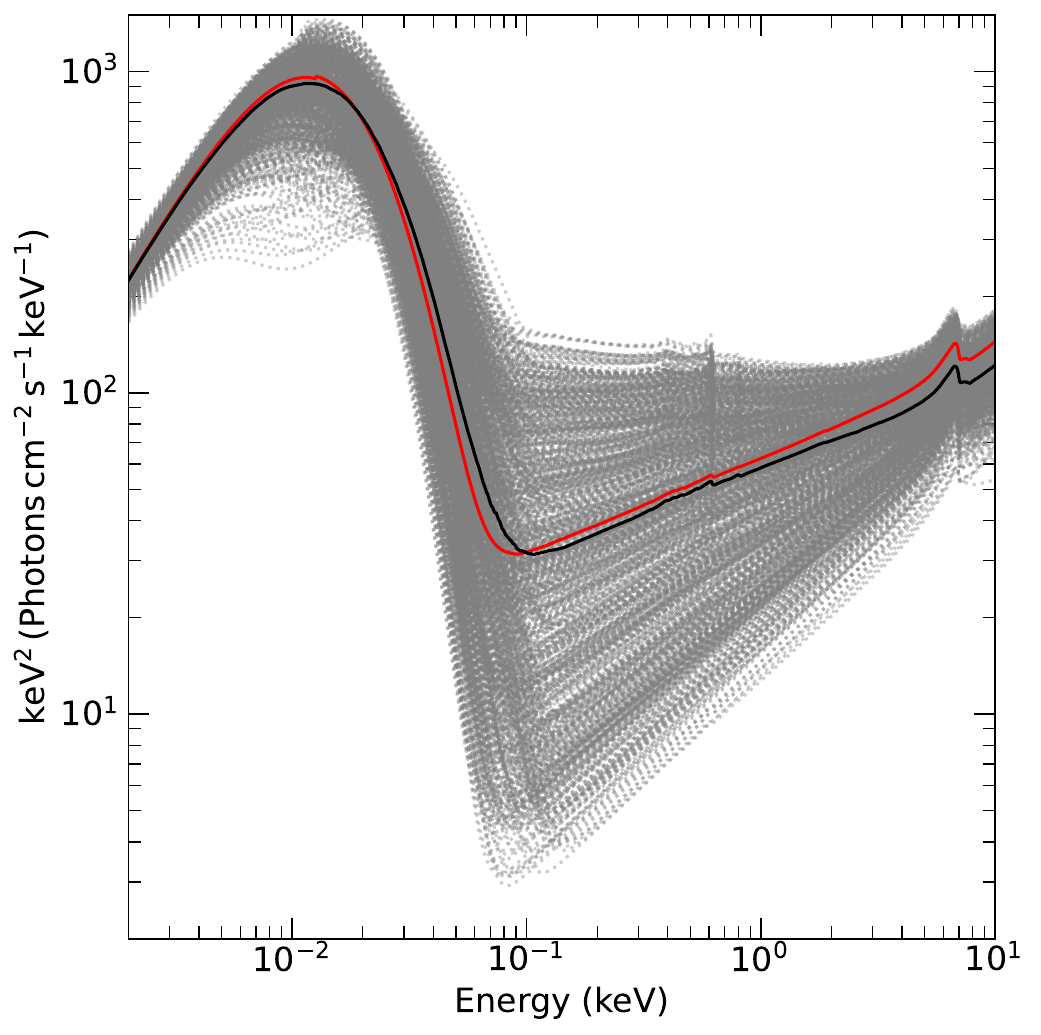}
    \caption{{Model SEDs estimated by choosing randomly 1000 values of $h$, $\Gamma$, and \ltransf\ (see Sect.\,\ref{sec:meanSED}). The individual realisations are shown as grey dotted lines. The black solid line correspond to the median of all 1000 SEDs. The red line corresponds to the SED estimated by considering the average values of $h$, $\Gamma$, and \ltransf.}}
    \label{fig:random_SED}
\end{figure}

\subsection{Comments on the physical meaning of the mean SED.}
\label{sec:meanSED}

{K23 presented the fits to the time-lag spectrum of NGC\,5548. These fits were performed by assuming $\mbh = 7\times 10^{7}\,M_\odot$ \citep{Horne2021} and an inclination angle of $40\degr$ \citep{Pancoast2014}. K23 showed that the model explains the observed time lags for a broad range of parameters. To reduce the degeneracy between these parameters, they considered a grid of $\spin, \fcol,$ and $\ltransf$, and fitted for the accretion rate and the corona height. For NGC\,5548, they considered as `best fit' only the cases which resulted in \mdot\ between 0.025 and 0.1, and that could also reproduce the observed X-ray luminosity.}

{The chosen range of \mdot\ is based on the fact that the accretion rate in this source is assumed to be $\sim 0.05$ of the Eddington limit \citep{Fausnaugh16}. This is based on the observed Eddington ratio $\lambda_{\rm Edd}=L_{\rm bol}/L_{\rm Edd}$, where $L_{\rm bol}$ and $L_{\rm Edd}$ are the bolometric and Eddington luminosity, respectively. However, accretion rate estimates which are based on observed $\lambda_{\rm Edd}$ values are rather uncertain, for various reasons. As K23 explained, the bolometric luminosity is usually  computed by measuring the observed luminosity in a given spectral band and then applying a bolometric correction factor. The observed luminosity may not be representative of the mean luminosity in this band, and the correction factor is quite uncertain. Furthermore, computation of $L_{\rm Edd}$ requires accurate knowledge of the BH mass, while converting from $L_{\rm bol}/L_{\rm Edd}$ to \mdot\ requires a knowledge of the inclination of the system. In addition, the bolometric luminosity estimates do not take into consideration intrinsic reddening due to dust that may be present in the host galaxy \citep[see e.g.][]{Gaskell2023}. For al these reasons, K23 considered a conservative uncertainty on the observed $\lambda_{\rm Edd}$ by a factor of $\sim 2$. For NGC\,5548 this translates into a range between 0.025 and 0.1, if $\lambda_{\rm Edd} = 0.05$ \citep{Fausnaugh16}.} 

{Based on this, K23 found that the model can fit the time-lag spectrum of NGC\,5548 for 5 different model parameter combinations:}

\begin{itemize}
    \item $\spin = 0, \fcol = 1.7, \ltransf = 0.5, h \geq 8.4 \,\rg$;\\
    \item $\spin = 0, \fcol = 2.5, \ltransf = 0.9, h \in [18.9, 28.5]\,\rg$;\\
    \item $\spin = 0.7, \fcol = 1.7, \ltransf = 0.5, h \geq 64 \,\rg$;\\
    \item $\spin = 0.998, \fcol = 2.5, \ltransf = 0.5, h \in [5.6, 28.5]\,\rg$;\\
    \item $\spin = 0.998, \fcol = 2.5, \ltransf = 0.9, h \in [18.9, 42.7]\,\rg$.
    
\end{itemize}

\noindent {As reported in Table\,2 of K23, the model fits for all the above parameter combinations are statistically acceptable, in the sense that $p_{\rm null}>0.01$, in all cases. For that reason, we refer to the above five models as the `best-fit' models, although they are not exactly `best model fits' in the traditional statistical sense. They provide a good fit to the time lags (i.e. $p_{\rm null}>0.01$), with accretion rates within the limits we mentioned above,  given the assumption that \spin, \fcol, and \ltransf, have the values listed above. Obviously, given the a priori determination of a few model parameters, the confidence regions on the  corona heights which are reported above are smaller than the case in which all the parameters are left free in the first place. We suspect that the model can still fit the time-lags even if we consider BH mass very different from the mass that is measured for this source and/or accretion rates outside the range we consider, but we believe that these models cannot be considered as reliable.}

In the following, we explain how we fitted the average SED using Eq.\,\ref{xspecmdel}, and the models listed above, in order to investigate whether any or all of them can explain simultaneously the time lags and the SED. We explain below why this should be the case, hence why this is a meaningful test for the model.

The flux from a compact X-ray corona can vary due to changes in radius, temperature, optical depth, height, or the BH accretion rate \citep[see for example Eq. 12 in][]{Zhang23}. Similarly to almost all AGN, the X--ray light curve of NGC\,5548 from the STORM campaign is highly variable, at all sampled time scales. These variations imply that one, or even all of the corona physical parameters change constantly with time. Even if it is variable, the accretion rate itself cannot explain the observed X--ray variations in AGN. The variability amplitude of the X--ray light curves (in NGC\,5548 and almost all other AGN) is of larger than the variability amplitude of the UV/optical light curves. Consequently, the mechanism that powers the corona should be variable and, somehow, at least one of the coronal parameters (radius, temperature,  optical depth, and height) should vary with time.

In this case, it is not straightforward to understand what the mean SED represents in such a variable source. To investigate this issue in the case of a variable X--ray source which illuminates the disc below, we produce {1000} optical/UV/X-ray spectra using \kynsed, and assuming a BH mass of $7\times 10^7\,M_\odot$, an inclination angle of $40\degr$, and an accretion rate of 0.05 the Eddington limit ({just like NGC\,5548}). {Furthermore, we consider an X--ray source whose height changes randomly between 4\,\rg\, and 40\,\rg, \ltransf\ changes between 0.5 and 0.9, and $\Gamma$ changing between 1.3 and 2.1 (the latter being indicative of either temperature and/or optical depth variations), by drawing these parameters from uniform distributions. The range of the parameter values we considered corresponds to the range of the parameters we find later on (see Sect.\,\ref{sec:bestfitres}), where we study the SED variability of NGC\,5548.} These spectra are plotted in Fig.\,\ref{fig:random_SED} with grey dotted lines. 

{The solid black line in the same figure shows the median of all the 1000 SEDs and the red solid line shows the SED when we assume the mean of the varying model parameters. This figure shows that, at least for the adopted range of parameter values (which is appropriate for the source we study), the mean SED is fully consistent with the SED of the mean of the variable parameters\footnote{We observe a slight difference between the median SED and the SED of the mean parameter values in the X-ray band, above $\sim 2$ keV. This is due to the large range of $\Gamma$ variations we consider, and is a problem with all power-law-like spectra. If the spectrum has a power-law-like shape and changes in spectral slope, then the mean spectrum will not be equal to the power law with the mean slope.}. The same applies for the case when the parameters are correlated, as long as they vary within the range that is observed for NGC\,5548 (see Sect.\,\ref{sec:correlation} for more details). Consequently, in the case of NGC\,5548, fitting the mean SED with \kynsed\ will provide best-fit parameter values which will be very close to the mean of the variable physical parameters (except, perhaps, with $\Gamma$)}. We note that the changes in the parameters of the system may also affect the time lags computed using the total light curves from multi-wavelength campaigns, as we discuss below in Sect.\,\ref{sec:lags}.

\subsection{The best-fit results}

\begin{figure*}
    \centering
    \includegraphics[width = 0.49\linewidth]{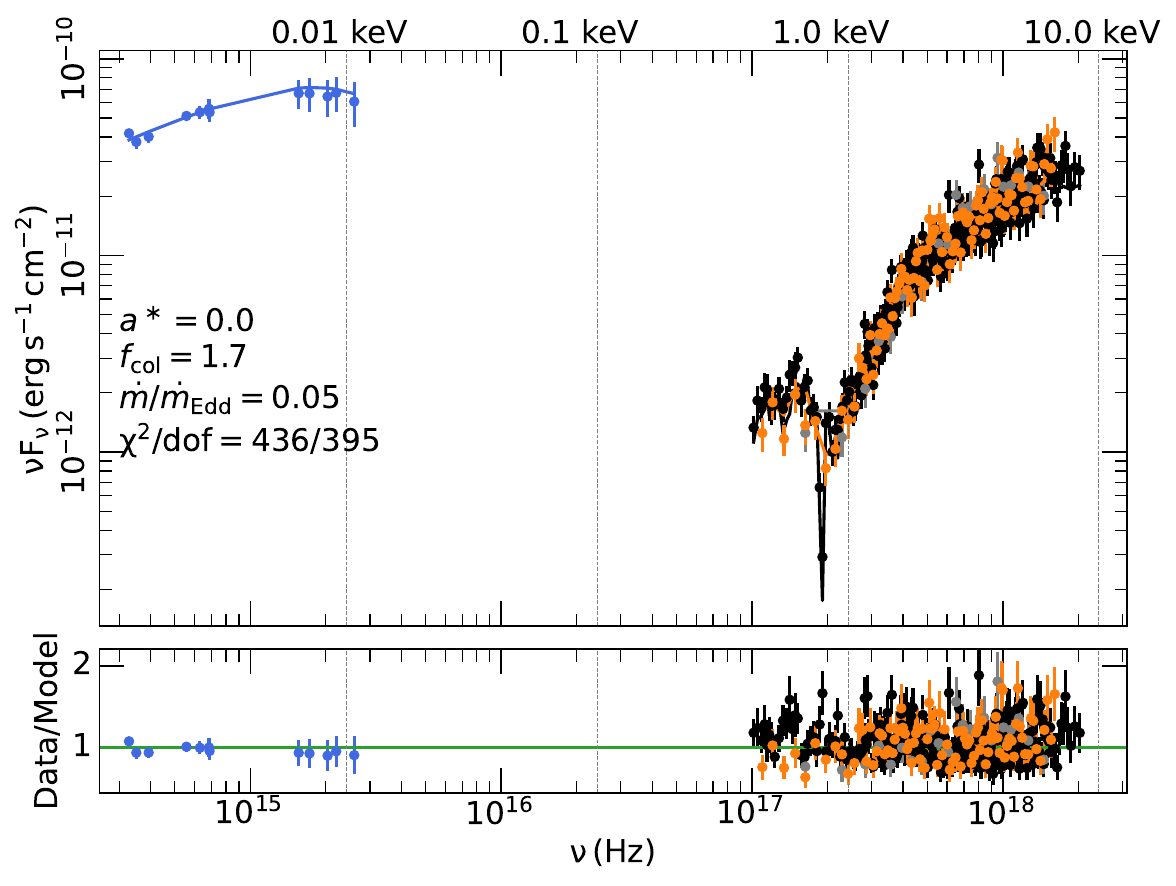}
    \includegraphics[width = 0.49\linewidth]{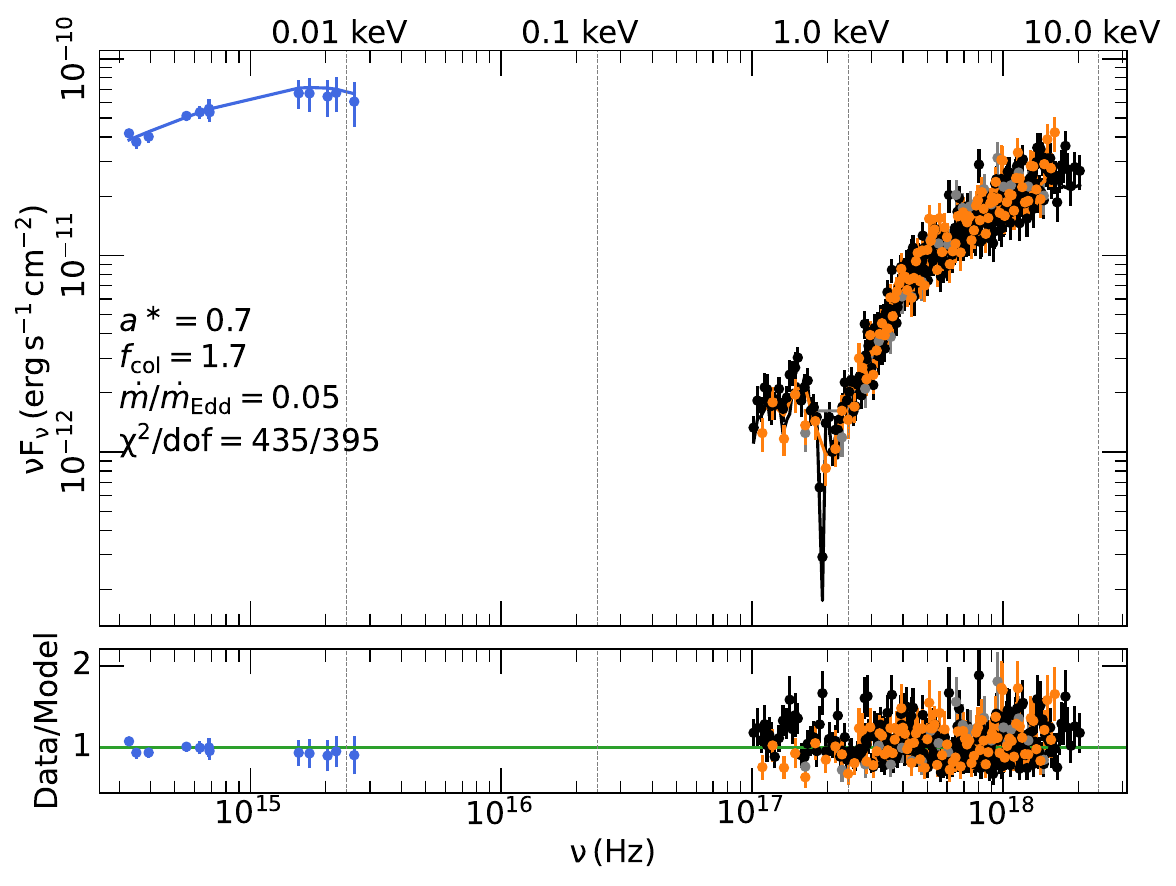}
    \caption{Time-averaged SED of NGC\,5548 fitted with Case 1 and Case 3 models, using the best-fit parameters listed in Table \ref{tab:bestpar_averageSED} (see \S\ref{sec:average spectrum} for details).
The blue data points correspond to the UV/optical data, the black, grey, and orange data points correspond to the \swift/XRT spectra extracted from three different time intervals that are consistent with the average X-ray flux. The bottom panels show the corresponding data-to-model ratios.}
    \label{fig:avsedfit}
\end{figure*}

\begin{table}
    \centering
        \caption{Best-fit $\chi^2$ values when fitting the average SED.}
    \label{tab:fit_averageSED}
    \begin{tabular}{lrrr}
    \hline \hline \\[-0.2cm]
   \mdot 	&	 0.025	&	0.05	&	0.1	\\	\hline \\[-0.2cm]
	&	\multicolumn{3}{c}{$\spin = 0$, $\fcol = 1.7$} 					\\[0.15cm]	
$\rm \chi^2/dof$	&	609/395	&	436/395	&	497/395	\\[0.15cm]
$p_{\rm null}$	&	$<0.01$	&	0.08	&	$<0.01$	\\[0.15cm]	\hline \\[-0.2cm]
	&	\multicolumn{3}{c}{$\spin = 0$, $\fcol = 2.5$} 					\\[0.15cm]	
$\rm \chi^2/dof$	&	786/395	&	503/395	&	443/395	\\[0.15cm]	
$p_{\rm null}$	&	$<0.01$	&	$<0.01$	&	0.05	\\[0.15cm]	\hline \\[-0.2cm]
	&	\multicolumn{3}{c}{$\spin = 0.7$, $\fcol = 1.7$} 					\\[0.15cm]	
$\rm \chi^2/dof$	&	629/395	&	435/395	&	437/395	\\[0.15cm]	
$p_{\rm null}$	&	$<0.01$	&	0.08	&	0.07	\\[0.15cm]	\hline \\[-0.2cm]
	&	\multicolumn{3}{c}{$\spin = 0.998$, $\fcol = 2.5$} 					\\[0.15cm]	
$\rm \chi^2/dof$	&	928/395	&	643/395	&	625/395	\\[0.15cm]	
$p_{\rm null}$	&	$<0.01$	&	$<0.01$	&	$<0.01$	\\	\hline
    \end{tabular}

\end{table}

The important physical parameters of \kynsed\ are: BH mass, \mbh, spin, \spin, accretion rate, \mdot, the ratio of the accretion power transferred to the corona over the total power given to the disc, \ltransf, the corona height, $h$, the X--ray spectrum parameters, $\Gamma$ and $E_{\rm cut}$, the disc inclination, and the colour correction factor, \fcol. Similarly to K23, we fix the BH mass at $7\times 10^7~\rm M_\odot$ \citep{Horne2021}, $E_{\rm cut}$ at 150\,keV \citep[in agreement with previous estimates by][]{Nicastro2000, Ursini15}, the inclination at 40\degr\ \citep[in agreement with the estimate by][]{Pancoast2014}. We also fix \mdot\ at three different values: 0.025, 0.05, and 0.1 {following the reasoning explained in K23 and in the previous section}. We then fit the average SED by considering the various combinations of \spin\ and \fcol\ that fit the time lags as presented above.  The free parameters are \ltransf, $h$, $\Gamma$, $E(B-V)_{\rm host}$, $N_{\rm H}$, and $\log \xi$. 

We note again that fixing \spin, \fcol, and \mdot\ will reduce the degeneracy that is present between some of the parameters, and will naturally lead to a decrease in the uncertainty on the free parameters (mainly \ltransf\ and $h$). Most probably, the SED can also be fitted well by different parameter combinations. But our objective is to investigate whether the same model parameter combinations are capable of explaining the SED as well as the time lags, since K23 fitted the time lags following the same approach. Obviously, our results should be considered as a sub-sample of a wider parameter space within the X-ray illuminated disc model that could fit the data.

We list in Table\,\ref{tab:fit_averageSED} the values of $\rm \chi^2/dof$ and $p_{\rm null}$ obtained for each of the combinations. There are four \spin, \fcol, and \mdot\, combinations that result in a statistically acceptable fit, namely: Case\,1: $\spin = 0, \fcol = 1.7, \mdot = 0.05$; Case\,2: $\spin = 0, \fcol =2.5, \mdot = 0.1$; Case\,3: $\spin = 0.7, \fcol = 1.7, \mdot = 0.05$; and Case\,4: $\spin = 0.7, \fcol = 1.7, \mdot = 0.1$. Although the combination of \spin=0.998 and \fcol=2.5 can provide good fits to the time lags, it cannot fit the average SED well, for any \mdot\ value within the range of 0.01--0.1 that we consider. 

Furthermore, two out of the four combinations which fit the SED well cannot be accepted as the resulting  best-fit $h$ and/or \ltransf\ values are not in agreement with the results obtained by fitting the time lags. For example, the Case\,2 best-fit results suggest $h>64\,\rg$ and $\ltransf \in [0.2,0.3]$ (these are the 90 per cent confidence limits
for a single parameter). These limits are in contradiction with the time lags best-fit results (for the same combination of \spin\ and \fcol), which implied $h \in [18.9, 28.5]\,\rg$ and $\ltransf = 0.9$. Similarly, according to the Case\,4 best-fit results, $h \in  [2.6,13.7]\,\rg$,  while $h\geq 64\,\rg$ according to the best-fit results to the time-lags. Thus, in the following we will adapt Cases\,1 and 3 as the only acceptable model parameters which can fit well both the time lags and the average SED. 

Table\,\ref{tab:bestpar_averageSED} lists the Case\,1 and Case\,3 best-fit parameters, and Fig.\,\ref{fig:avsedfit} shows the respective best-fit models to the average SED. We note that, except for the spin and the height, all the parameters are consistent between the two cases. The last row in Table\,\ref{tab:bestpar_averageSED} list the corona radius ($R_{\rm C}$) for both models. \kynsed\ computes a posteriori the corona radius for a spherical geometry, based on the best-fit model parameters, assuming conservation of energy and photon number. All the assumptions and the method of calculating $R_{\rm C}$ are detailed in \cite{Dovciak2016} and D22. A sanity check of the validity of the two model fits is that the corona should be outside the event horizon of the BH. Consequently, $R_{\rm C}$ should be smaller than the difference between the corona height and the event horizon radius ($R_{\rm H}$). For both cases, the corona is above the event horizon.

\begin{table}
    \centering
    \caption{Case 1 and Case 3, best-fit parameter values obtained by fitting the average SED of NGC\,5548.}
    \label{tab:bestpar_averageSED}
    \begin{tabular}{lll}
    \hline \hline \\[-0.2cm]
    Parameter & Case\,1& Case\,3 \\ \hline \\[-0.2cm]
    \spin	&	0		&	0.7	\\[0.15cm]	
\fcol	&	1.7		&	1.7	\\[0.15cm]	
\mdot	&	0.05		&	0.05	\\[0.15cm]	
$E(B-V)_{\rm host}$	&	$0.14_{-0.01}^{+0.02}$		&	$0.14_{-0.02}^{+0.01}$	\\[0.15cm]	
\ltransf	&	$0.55_{-0.01}^{+0.02}$		&	$0.54_{-0.01}^{+0.02}$	\\[0.15cm]	
$h\,(\rg)$	&	$31.4_{-9.1}^{+10.4}$		&	$79.5_{-10.5}^{+11.5}$	\\[0.15cm]	
$\Gamma$	&	$1.83_{-0.04}^{+0.06}$		&	$1.81_{-0.06}^{+0.07}$	\\[0.15cm]	
$N_{\rm H}\,(10^{22}\rm cm^{-2})$	&	$2.54_{-0.25}^{+0.28}$		&	$2.49_{-0.24}^{+0.26}$	\\[0.15cm]	
$\log \left( \xi/\rm erg\,cm\,s^{-1}\right)$	&	$1.37_{-0.08}^{+0.10}$		&	$1.36_{-0.07}^{+0.09}$	\\ \hline \\[-0.2cm]	
$R_{\rm C}\,(\rg)$	&	23		&	31	\\[0.15cm]	\hline
    \end{tabular}
    
\end{table}

\section{Modelling the complex X--ray/UV/optical SED variability in NGC 5548}
\label{sec:variableSED}

\begin{figure*}
    \centering
    \includegraphics[width = 0.9\linewidth]{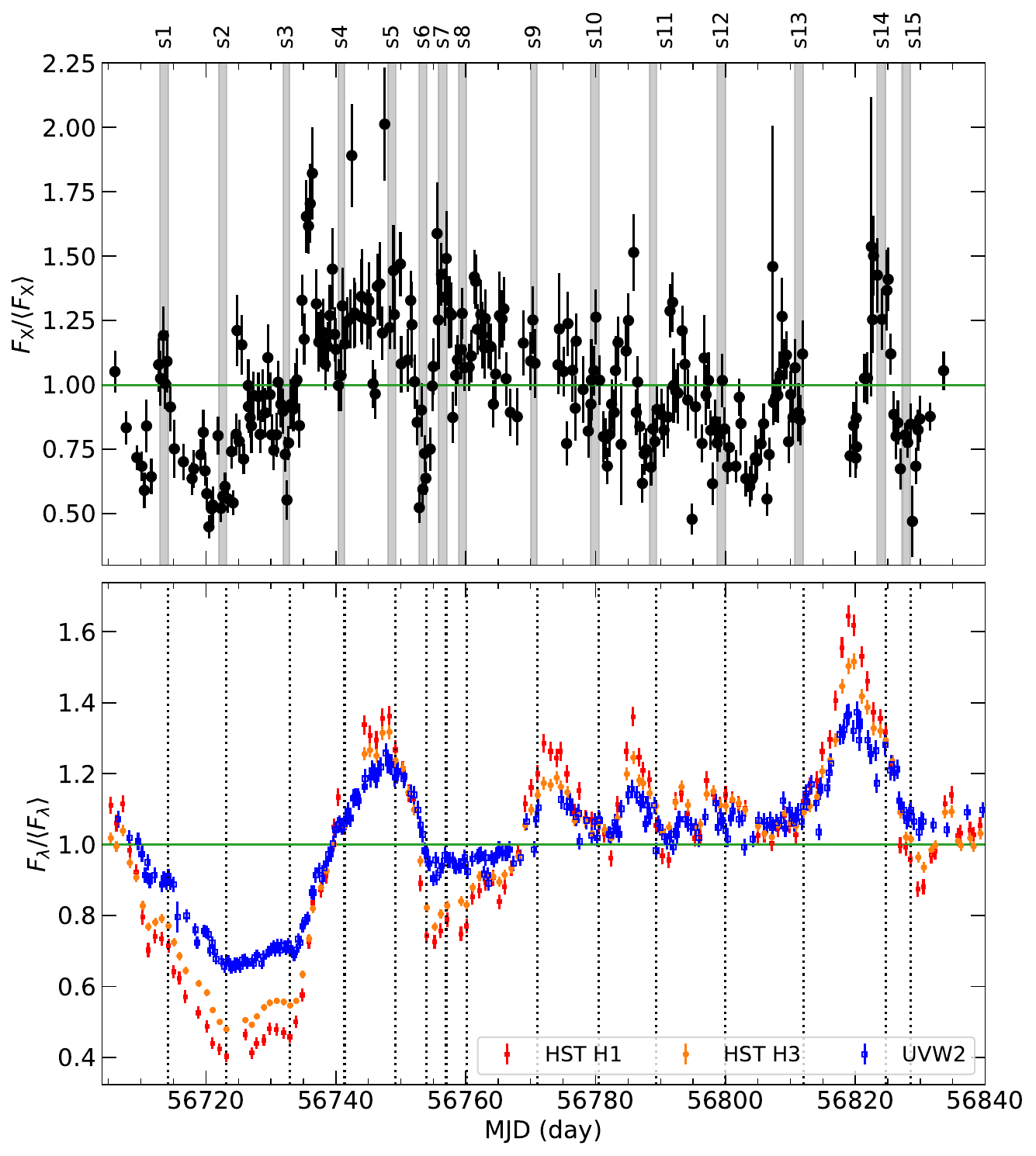}\\
    \caption{Mulit-wavelength light curves of NGC\,5548. Top and bottom panels: The observed X-ray and UV light curves normalised to their mean (H1, H3, and UVW2 bands are plotted in red, orange, and blue points, respectively). The vertical dotted lines indicate the times at which we selected the UV data points. The grey shaded areas show the corresponding time intervals over which we integrated the X-ray spectrum (See text for details).}
    \label{fig:lightcurve}
\end{figure*}

Figure~\ref{fig:lightcurve} shows the \swift/XRT light-curve of NGC\,5548 \citep[taken from][]{Panagiotou20}, the \hst\ light curves at 1158\,\AA\ and 1479\,\AA\ (hereafter H1 and H3) and the \swift/UVOT light curve in the UVW2 band, centred at 1928\,\AA \citep[taken from][for the UVOT and \hst\ data, respectively]{Edelson15, Fausnaugh16}. The light curves are normalised to their corresponding mean, and they show clearly the complex variability of the source. For example, we can see mini-flares and dips in X--rays which are absent in the UV bands, the X--ray flux decreases (overall) in the period between $\sim 56760-56810$ MJD, while the UV flux appears to increase, etc. Even within the UV bands, there are differences in the variability amplitude, which imply significant UV spectral variability with the spectrum becoming bluer with increasing flux. Any physical model should be able to explain these complex spectral variability patterns. 

The vertical dotted lines and the shaded areas (in the bottom and top panels in Fig.\,\ref{fig:lightcurve}) indicate 15 times (in the UV) and 15 X--ray light curve segments (s1-s15), which sample characteristic features in the X--ray light curve such as maxima (e.g. s1, s7), minima (e.g. s6, s15), or large amplitude variations (s8-s12), which do not necessarily correspond to similar features/trends in the UV light curves. Our objective is to extract the UV/X--ray SEDs during these times, fit them with \kynsed\, and investigate whether X--ray reverberation can explain the complex X--ray/UV variability in this source, the origin of the X--ray variability as well as the reason for the rather poor X--ray/UV correlation in this source, that is to say why the X--ray and the UV fluxes do not appear to vary in phase, at all times. 

\subsection{Constructing the time-resolved SEDs}

\begin{figure}
    \centering
    \includegraphics[width = 1.0\linewidth]{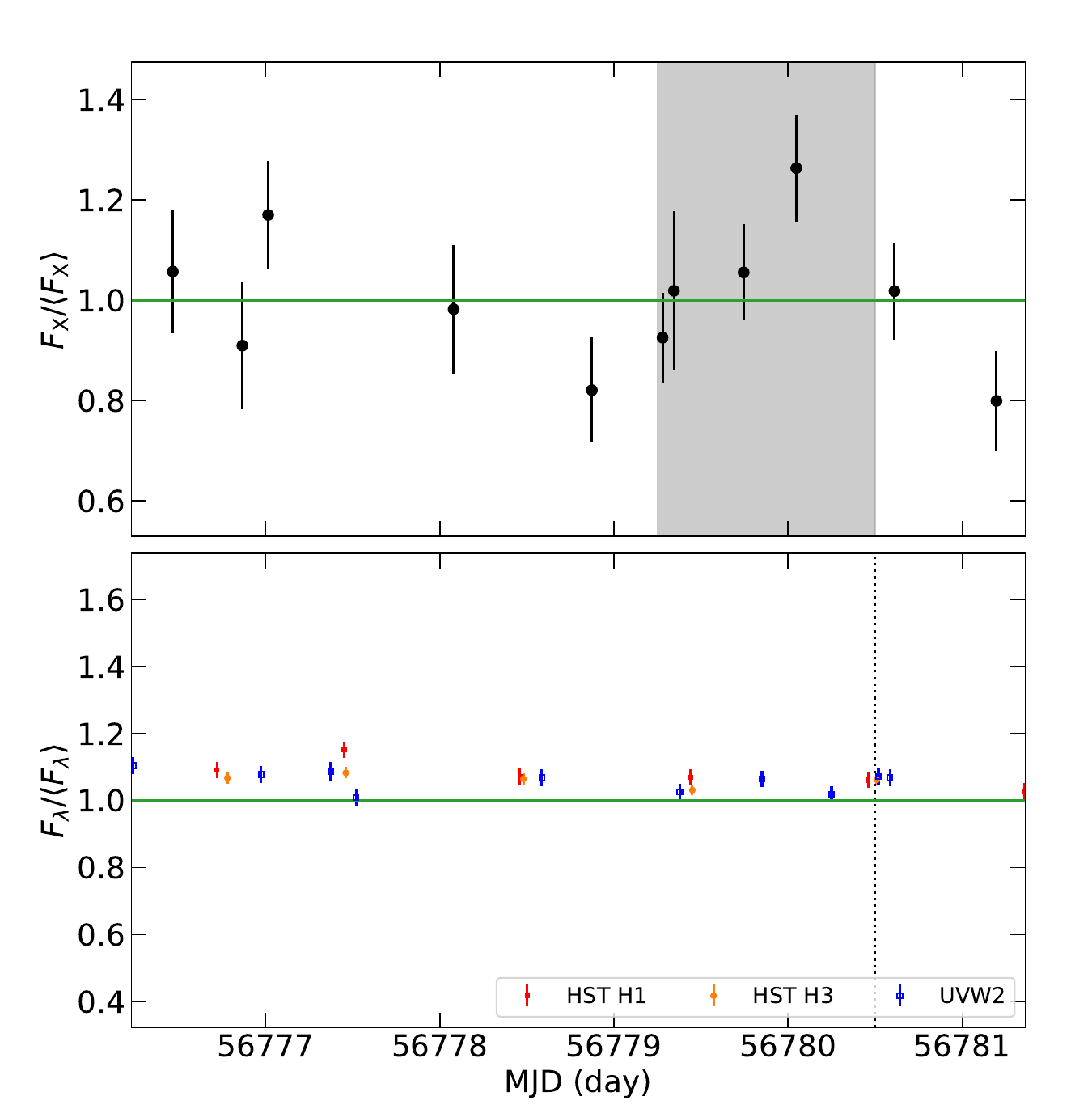}
    \caption{Zoom-in on the time segment s10 as an example of the data selection. The top and bottom panels are similar to the ones shown in Fig.\,\ref{fig:lightcurve}.}
    \label{fig:segment}
\end{figure}
If X--rays illuminate the accretion disc and produce the variable UV/optical emission, we can{\it not} simply use data from simultaneous UV/X--ray observations to construct the SED. For example, Fig.\,\ref{fig:segment} shows a small part of the X--ray and UV light curves around segment s10, as labelled in Fig.\,\ref{fig:lightcurve}. The vertical dotted line in the bottom panel of Fig.\,\ref{fig:segment} indicates the H1, H3 and UVW2 points we use for the s10 SED. There is an \swift/XRT observation just after the dotted line but we can{\it not} consider the X--ray spectrum from this observation to construct the X--ray/UV SED. 

In the case of X--ray thermal reverberation, the variable UV emission at any time does not depend on the X--rays that were emitted at that time. Instead, this emission should be equal to the convolution of the X--rays with the disc response function at all times {\it before} the UV observations took place (see Eq. (4) in K21, for example), and it is the width of the response function that determines how far in the past we should consider this convolution. According to Fig.\,8 in K21a, the disc response function at $\sim 1200$ \AA\ in the case of M$_{\rm BH}=5\times 10^7$ M$_{\odot}$ is $\sim 10-15$ time smaller than its peak $\sim 1$ day  after the X--rays were emitted (both for \spin=0 and 1). In the case of the UVW2 band, the response function is $\sim 10$ time smaller than its peak at approximately $\sim 2$ days after the X--ray emission. Consequently, the X--ray spectrum in the s10 SED should be the average X--ray spectrum over a period of $\sim 1-2$ days before the UV data. 

We therefore extract the spectrum from all the X--ray observations that were taken within $1.3$\,days {\it before} the UV observations (we use 1.3 days as a compromise between 1 and 2 days). This time interval is indicated by the grey area in the upper panel in Fig.\,\ref{fig:segment}. In fact, all the grey-shaded areas in the top panel in Fig.\,\ref{fig:lightcurve} indicate the 1.3 day intervals we consider to extract the X--ray spectrum for each UV/X-ray SED. There are typically $3-4$ \swift/XRT observations in such a period for the SED of all segments.  

We cannot add longer wavelengths in the SED because the width of the disc response function increases with increasing wavelength. For example, Fig.\,8 in K21a shows that the width of the response function is at least $\sim 10$ days in the $g'$ band. Consequently, if we wish to include longer wavelengths in the SED, then we should  extract the X-ray spectrum using observations over a period longer than 1--2 days. But then, we must construct different X--ray spectra for the UV and the optical SEDs at each time, and the model fitting of such data sets would have been very challenging. 

We extract the X–ray spectra of each interval using the automatic \swift/XRT data products generator\footnote{\url{https://www.swift.ac.uk/user_objects/}} \citep{Evans09}. We consider the $0.4-8$\,keV band in each spectrum, and grouped the spectra to have at least 25~counts per bin. For the UV/optical data, we use the {\tt FTOOLS} \citep{Ftools2014} command {\tt ftflx2xsp}\footnote{\url{https://www.heasarc.gsfc.nasa.gov/lheasoft/ftools/headas/ftflx2xsp.html}} to create spectral files in the UV/optical range in a format that is compatible with {\tt XSPEC}, which we use to model the SED as described in the following section. In the following, all uncertainties are estimated for a change of fit statistic by $\Delta \chi^2 = 2.706$ for a single parameter. 

\subsection{Spectral fitting of the variable SEDs}
\label{sec:fitting}
\begin{figure}
    \centering
    \includegraphics[width = 1\linewidth]{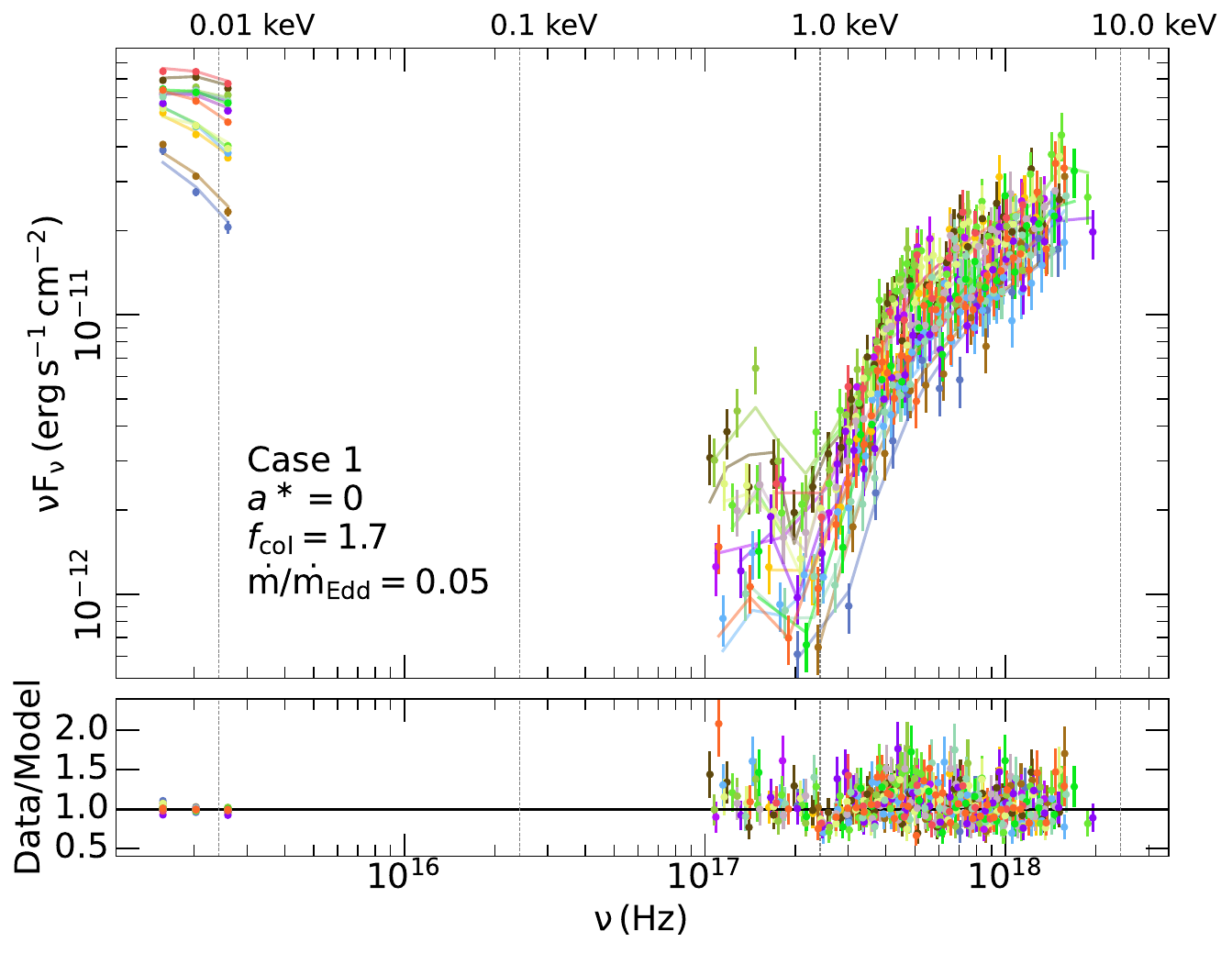}
    \caption{Best-fit to all 15 SEDs for the Case\,1 model. The corresponding ratios (data/model) are plotted in the bottom panel.}
    \label{fig:SEDbestfit}
\end{figure}

We fitted all the 15 SEDs with the model defined by Eq.\,\ref{xspecmdel} twice. First, we fixed spin and \fcol, to 0 and 1.7, and then to 0.7 and 1.7. These are the Case 1 and Case 3 model values, which fit well both the average SED as well the time-lags spectrum of the source. Both times, we assume the BH mass, inclination and $E_{cut}$ values that we use when fitting the average SED, and we fix the covering fraction of the ionised absorber to 1. Each time, we fix the accretion rate, and the host galaxy extinction to the best-fitting values listed in Table\,\ref{tab:fit_averageSED}. We then fit separately each of the 15 SEDs letting free the height, \ltransf, the photon index, the column density and the ionisation parameter of the ionised absorber.

The majority of the spectra resulted in statistically unacceptable fits when assuming the Case\,3 model (i.e. $p_{null}$ is smaller than 0.01 in most cases). On the contrary, all 15 SEDs  could be well fitted assuming Case\,1, model. The only exception was the s2 SED, in which case  $\chi^2$/dof=32.8/13, $p_{null}=0.002$. Adding a systematic error on the model for this fit of 3\%, using the {\tt systematic} command in {\tt XSPEC}, improved the fit significantly (rendering the fit statistically acceptable). We also compute the corona radius, $R_{\rm C}$, when fitting each SED. In all cases, the radius was physically accepted, i.e. it was smaller than the corona height minus the BH horizon, except for the s8 and s9 SEDs. For these two spectra, the best-fit $\Gamma$ exceeded 2.3, which in turn results in $R_{\rm C}$ values larger than the height of the corona. For that reason, we set an upper limit of $\Gamma=1.9$ for these two spectra (this is consistent with the upper uncertainty we get on $\Gamma$ by fitting the time-averaged SED). By doing so, the fit remains statistically accepted and the value of $R_{\rm C}$ become physically accepted as well. Figure\,\ref{fig:SEDbestfit} shows the 15 SEDs together with the corresponding best-fitting model.

\subsection{Best-fit results}
\label{sec:bestfitres}

\begin{figure}
    \centering
    \includegraphics[width = 0.95\linewidth]{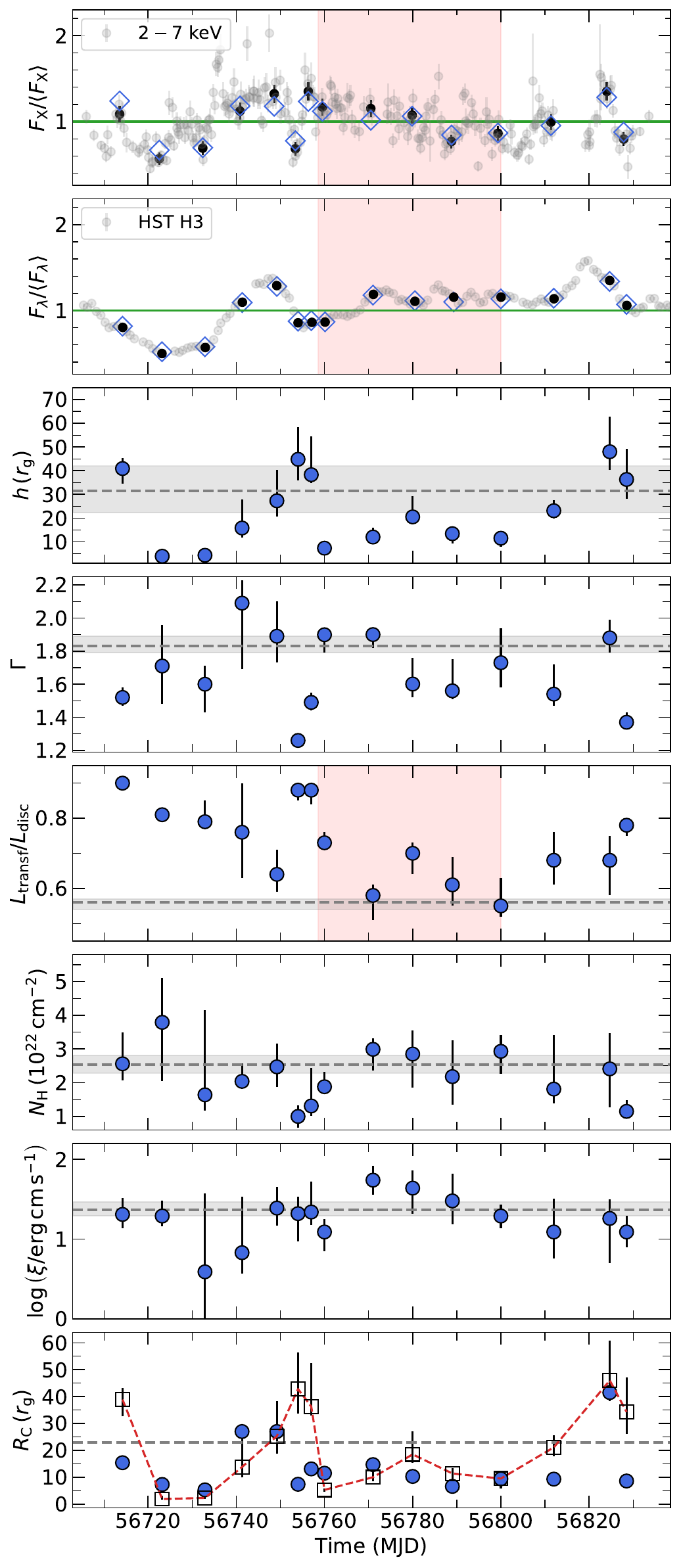}
    \caption{Evolution of the best-fit parameters as fuction of time.Top two panels show the observed and the best-fit model-predicted X-ray and UV light curves (top and bottom panels, respectively). Grey circles show the mean count rate from all observations, while the black circles indicate the average count rate in each of the 15 time segments we computed the SEDs. Blue diamonds indicate the best-fit flux in each band, during the same segments. All fluxes and the black circles count rates are normalised to their respective mean. The bottom six panels show the evolution of the best-fit parameters as a function of time. The horizontal grey dashed lines and shaded areas correspond to the best-fit parameters (and the corresponding uncertainty) obtained by fitting the time-averaged SED in Sect.\,\ref{sec:average spectrum}. Blue circles in the bottom panel show the corona radius, and the empty black squares indicate the  difference between the height of the corona and the event horizon radius (errors on these points take into account only the error on the height). {The red shaded area highlights the s8-s12 period in which \ltransf\ is the main driver of the variability (see Sect.\,\ref{sec:correlation} for more details).}}
    \label{fig:param_lc}
\end{figure}

The best-fit parameter values for all the SED model fits are listed in Table\,\ref{tab:results}. Using these values, we compute the model X--ray flux in the $2-7$\,keV band, as well as the UV flux in the bands we consider. Grey circles in {the top two panels of} Fig.~\ref{fig:param_lc} correspond to the observed mean count rate during each observation in the X--ray and in the H3 band (first and second panels, respectively). Black circles in both panels indicate the count rates during the times chosen to construct the 15 X--ray/UV SEDs, normalised to their mean. The open diamonds in both panels indicate the best-fit model luminosity in each band, normalised to the respective mean values. The agreement between the model and the observed fluxes is very good. Just like the model/data ratios in Fig.\,\ref{fig:SEDbestfit}, this figure demonstrates clearly the quality of the model fits to the data. 

Figure~\ref{fig:param_lc} shows that the model and the observed UV fluxes, plotted in the {second panel in Fig.\,\ref{fig:param_lc}}, agree at the $\sim 1.5\%$ level, on average. Since we fitted the SEDs by keeping \mdot\ constant, this implies that X--ray thermal reverberation could account for {\it all} the observed UV variations in NGC\,5548.  This result is even more impressive given the fact that the correlation between the X--rays and the UV is rather weak. As we have already discussed, this fact raised the question of whether the X-rays are the main driver of the variations seen in the UV/optical or not. Our results show that, despite the fact that the two light curves appear to be weakly correlated, it is entirely possible that all the UV variations are caused by the X--rays which illuminate the disc. The reason for the less than perfect correlation between the two light curves can be understood, to some extent, when we investigate the variability of the best-fit model parameters. 

Figure \ref{fig:param_lc} shows how the best-fit parameters vary as a function of time. The horizontal dashed lines and the grey shaded areas around them indicate the best-fit values we get when we fitted the mean SED, and their 1$\sigma$ confidence limit, respectively. The column density and ionisation parameter of the ionised absorber do not appear to vary significantly. They remain roughly constant close to the best-fit values we get when we fitted the mean SED. On the other hand, the corona parameters, i.e \ltransf, $\Gamma$, and height, are highly variable. The height of the corona changes significantly in the first 8 segments, and then remains roughly constant between $\sim 10-20$~\rg\ before increasing in the last two segments above $30\,\rg$. In particular, the height appears to determine the X--ray and UV variability during the period between s1 and s5 SEDs. During this period, the X--ray and UV flux vary in the same way (as shown by the first 5 open diamonds in {the top two panels of Fig.\,\ref{fig:param_lc}}). The third panel in Fig.\,\ref{fig:param_lc} shows that the height varies in exactly the same way. As the height decreases (from s1 to s2) the X--ray and the UV flux decrease as well. This is because as the height decreases X--rays are strongly lensed towards the BH, the X-rays detected by the observer decrease, and so are the X--rays illuminating the disc, hence producing less of the variable UV flux. As the height increases (for s3 to s4 and s5). then both the X-rays emitted towards the observer and towards the disc increase as well. 

We also see \ltransf\ variations as well. This parameters also influences significantly the observed X--ray and UV variations. For example, in the period between s8 to s12, the X--ray flux systematically decreases, while the UV flux increases. This apparent contradictory behaviour can be explained by the fact that \ltransf\ decreases during the same period (see Fig.\,\ref{fig:param_lc}, in the period between 56760--56780 MJD). For that reason the X--ray flux decreases as well but, since the power that is transferred to the corona is taken by the accretion process, as \ltransf\ decreases, the disc luminosity increases, and hence the UV flux as well. Therefore, although the two light curves during the period between the s8 and s12 segments appear to vary in the opposite directions, this can be explained, and is entirely consistent, with the X--ray, disc reverberation. Within the context of our model, as long as the power transferred to the corona is  part of the accretion power released in the disc, and it is variable, then the X--ray and the UV flux variations can be even anti-correlated.

The last panel in Fig.\,\ref{fig:param_lc} shows the radius of the corona during the full period, computed as explained in D22. In almost all cases the radius is smaller than or equal (within the errors) to the height of the corona (minus the horizon radius), as it should be. The radius does not vary a lot, and most of the time sit is $\sim 10$\,\rg, although it reduces to less than 5\,\rg, during s2 and s3 (when the height is small as well), and in a couple of cases (s4, s5 and s15) it may be up to 20 or even 40\,\rg. 

\subsection{Comparison between the best-fit results to the mean and to the time variable SEDs}

\begin{figure}
    \centering
    \includegraphics[width = 1\linewidth]{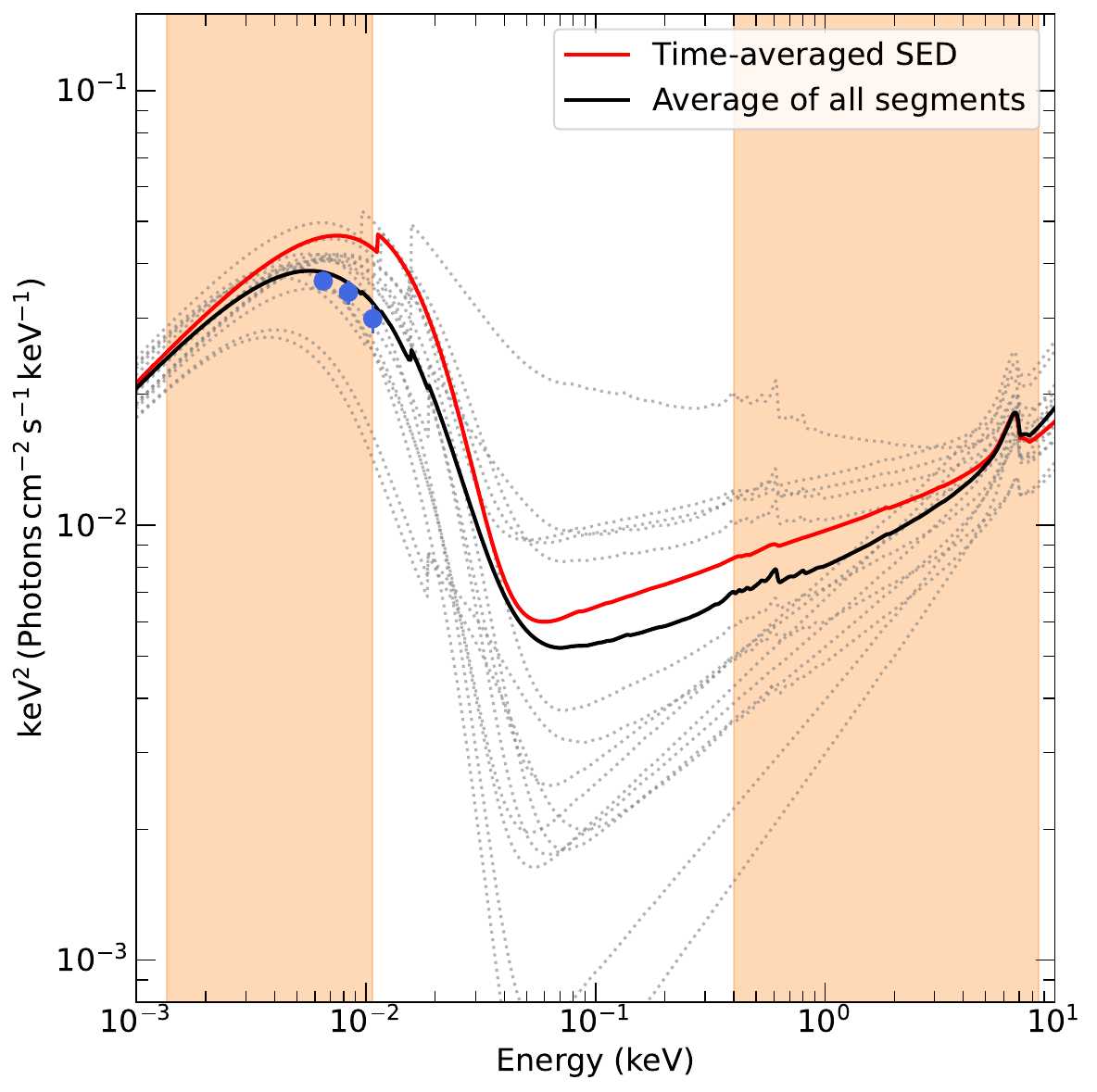}
    \caption{Best-fit model SEDs to the 15 spectra of NGC\,5548 (dotted grey lines). The solid black line shows the average of the 15 SEDs. The solid red line corresponds to the best-fit model to the time-averaged SED. Blue circles indicate the average flux of the 15 segments in the UV bands. The orange shaded areas highlight the energy ranges used to fit the time-averaged SED.}
    \label{fig:average_modelSED}
\end{figure}


Figure \ref{fig:param_lc} shows that the best-fit log($\xi)$ and $N_{H}$ values when we fit the time variable SEDs and the mean SED are in agreement. But this is not the case with the other model parameters. The height of the corona during the 15 segments is systematically smaller than the height we measure when we fit the mean SED. The opposite is true with \ltransf. The \ltransf\ we measure when we fit the average SED is systematically lower than the best-fit \ltransf\ values we get when we fit the individual SEDs. The disagreement between the mean of the model-parameters of the 15 SEDs and the model parameters of the (overall) mean SED is because the segments we choose to study the spectral variability of the source were not chosen so that their mean flux would be equal to the overall mean of the light curves. It turns out that the mean flux of these segments is smaller than the mean of the whole light curves, and this can explain the differences we see.

The grey dotted lines in Fig.\,\ref{fig:average_modelSED} indicate the best-fit \kynsed\ models to the 15 variable SEDs. The solid black line shows the mean of these 15 models. Of course, the mean of the 15 best-fit models should give a good fit to the mean observed spectrum of the 15 segments. Indeed, the three blue points in Fig.\,\ref{fig:average_modelSED} show the mean H1, H3 and UVW2 flux of the UV observations in the 15 segments, and they are in perfect agreement with the black solid line. The solid red line in the same figure shows the best \kynsed\ model-fit to the mean SED. The mean spectrum of the individual SEDs is lower than the overall mean SED. Consequently, it is not necessary that the mean of the parameter values from the best-fits to the 15 SEDs is equal to the respective model parameters we derive from the model fit to the average SED. 

At first, it seems rather strange that the mean \ltransf\ during the 15 segments is larger than the \ltransf\ we determined when we fitted the average SED, while the mean X--ray flux during the 15 segments is lower than the overall mean. We would expect the opposite, but, at the same time, the mean corona height is quite smaller than the height we measure when we fitted the mean SED. Therefore, although the power transferred to the corona during the 15 segments is larger than the average power, a larger fraction of the X--rays are emitted away from the observer, towards the BH, so that the observed X--ray flux is smaller than average.

\section{Summary and discussion}
\label{sec:discussion}

This is the seventh paper in which we study the X-ray and UV/optical variability of NGC\,5548, using the data originally collected during the STORM campaign, assuming X-ray illuminated accretion disc. In \cite{Kammoun19lags, Kammoun21b} we fitted the UV/optical continuum time lags assuming that the energy powering the X--ray source was not part of the power released in the disc through accretion. Assuming the same scenario, we were also able to explain the UV/optical power spectral density (PSD) of the source in \cite{Panagiotou20, Panagiotou22}. Later on, D22 presented {\tt KYNSED}, which can be used to model the broadband SED of AGN assuming X-ray illumination and by allowing for two scenarios for powering the corona: either the power is external to the accretion disc, or the corona is powered via the accretion process. D22 used the latter scenario to model the broadband time-averaged SED of NGC\,5548. More recently, K23 modelled the time-lags of the source assuming the same scenario. In this work, we use the best-fit results obtained by K23 to model both the time-averaged and the variable SEDs of the source. In the latter case, we consider 15 segments of the X-ray and UV bands (at wavelengths shorter than $\sim 2000$\,\AA) and we model the complex, X--ray/UV spectral variability of the source.

The scenario of an X-ray illuminated accretion disc can explain the time lags and the time-averaged SED of NGC\,5548 with the same model parameters. In this case, two values of the BH spin are acceptable (0 and 0.7), assuming $\mdot = 0.05$. Both spin values result in comparable values of $\Gamma \sim 1.8$, $\ltransf \sim 0.55$. However, the lower spin value resulted in $h \sim 31\,\rg$, while the higher spin value resulted in $h \sim 80\,\rg$. To the best of our knowledge, our model is the only model that can explain both the spectral as well as the timing properties of NGC\,5548. Furthermore, this model can even explain the time-variable X-ray/UV SEDs of the source, assuming a constant accretion rate. The fact that the X-rays are the only driver of the UV/optical variability can resolve problems associated with the fast UV/optical variations observed in AGN, and could not be explained by the accretion disc (slow) characteristic time scales. 

We find that only the zero BH spin case can explain the variable SED. The spectral variability can be thus explained by variations in the corona height (between $\sim 4\,\rg$ and $50\,\rg$), \ltransf\ (between $\sim 0.55$ and 0.9), and $\Gamma$ (between $\sim 1.25$ and 2.1). Since these model parameters affect the observed flux in various bands in different ways, the model can account for the weak correlation that is observed between the X-rays and the UV/optical light curves \citep{Fausnaugh16}. 

\subsection{The origin of the X--ray variability and the weak X--ray/UV correlation}
\label{sec:correlation}

\begin{table}
    \centering
    \tiny
    \caption{Best-fit slopes obtained by fitting a straight line, in linear or logarithmic scale, for various combinations of the best-fit model parameters. 
     {Values in parenthesis below the best-fit slopes indicate the significance of the hypothesis that the slope is different than zero. In the last column, we list the Spearman's rank correlation coefficient and the null hypothesis probability (values in parenthesis).}}
    \label{tab:linefit}
    \begin{tabular}{llll}
    \hline \hline \\[-0.15cm]
    Parameters & Space & Slope & $\rho_{\rm S}$\\ \hline \\[-0.15cm]

    $h\,(r_{\rm g})$ vs $L_{\rm transf}/L_{\rm disc}$ & linear & $56.92 \pm 22.94$ & 0.27 \\[0.1cm]
     &  & ($2.5 \sigma$) & (0.33) \\[0.1cm]
    $\Gamma$ vs $L_{\rm transf}/L_{\rm disc}$ & linear & $-1.00 \pm 0.35$ & -0.56 \\[0.1cm]
     &  & ($2.8 \sigma$) & (0.03) \\[0.1cm]
    $F_{\rm X}/\langle F_{\rm X} \rangle$ vs $L_{\rm transf}/L_{\rm disc}$ & linear & $0.04 \pm 0.46$ & 0.00 \\[0.1cm]
     &  & ($0.1 \sigma$) & (1.00) \\[0.1cm]
    $F_{\rm UV}/\langle F_{\rm UV} \rangle$ vs $L_{\rm transf}/L_{\rm disc}$ & linear & $-1.41 \pm 0.30$ & -0.79 \\[0.1cm]
     &  & ($4.6 \sigma$) & (0.0004) \\[0.1cm]
    $\Gamma$ vs $h\,(r_{\rm g})$ & linear & $-0.006 \pm 0.004$  & -0.46 \\[0.1cm]
     &  & ($1.6 \sigma$ )& (0.08) \\[0.1cm]
    $F_{\rm X}/\langle F_{\rm X} \rangle$ vs $h\,(r_{\rm g})$ & log & $0.15 \pm 0.06$  & 0.57 \\[0.1cm]
     &  & ($2.5 \sigma$)  & (0.03) \\[0.1cm]
    $F_{\rm UV}/\langle F_{\rm UV} \rangle$ vs $h\,(r_{\rm g})$ & log & $0.19 \pm 0.08$  & 0.31 \\[0.1cm]
     &  & ($2.4 \sigma$)  & (0.26) \\[0.1cm]
    $F_{\rm X}/\langle F_{\rm X} \rangle$ vs $\Gamma$ & log & $0.62 \pm 0.21$  & 0.26 \\[0.1cm]
     &  & ($3.0 \sigma$)  & (0.35) \\[0.1cm]
    $F_{\rm UV}/\langle F_{\rm UV} \rangle$ vs $\Gamma$ & linear & $0.33 \pm 0.16$ & 0.37 \\[0.1cm]
     &  & ($2.1 \sigma$) & (0.18) \\[0.1cm]
    $F_{\rm UV}/\langle F_{\rm UV} \rangle$ vs $F_{\rm X}/\langle F_{\rm X} \rangle$ & linear & $1.13 \pm 0.11$ & 0.38 \\[0.1cm]
     &  & ($9.9 \sigma$) & (0.16) \\[0.1cm]
     \hline
    \end{tabular}
    
\end{table}

\begin{figure}
    \centering
    \includegraphics[width = 1\linewidth]{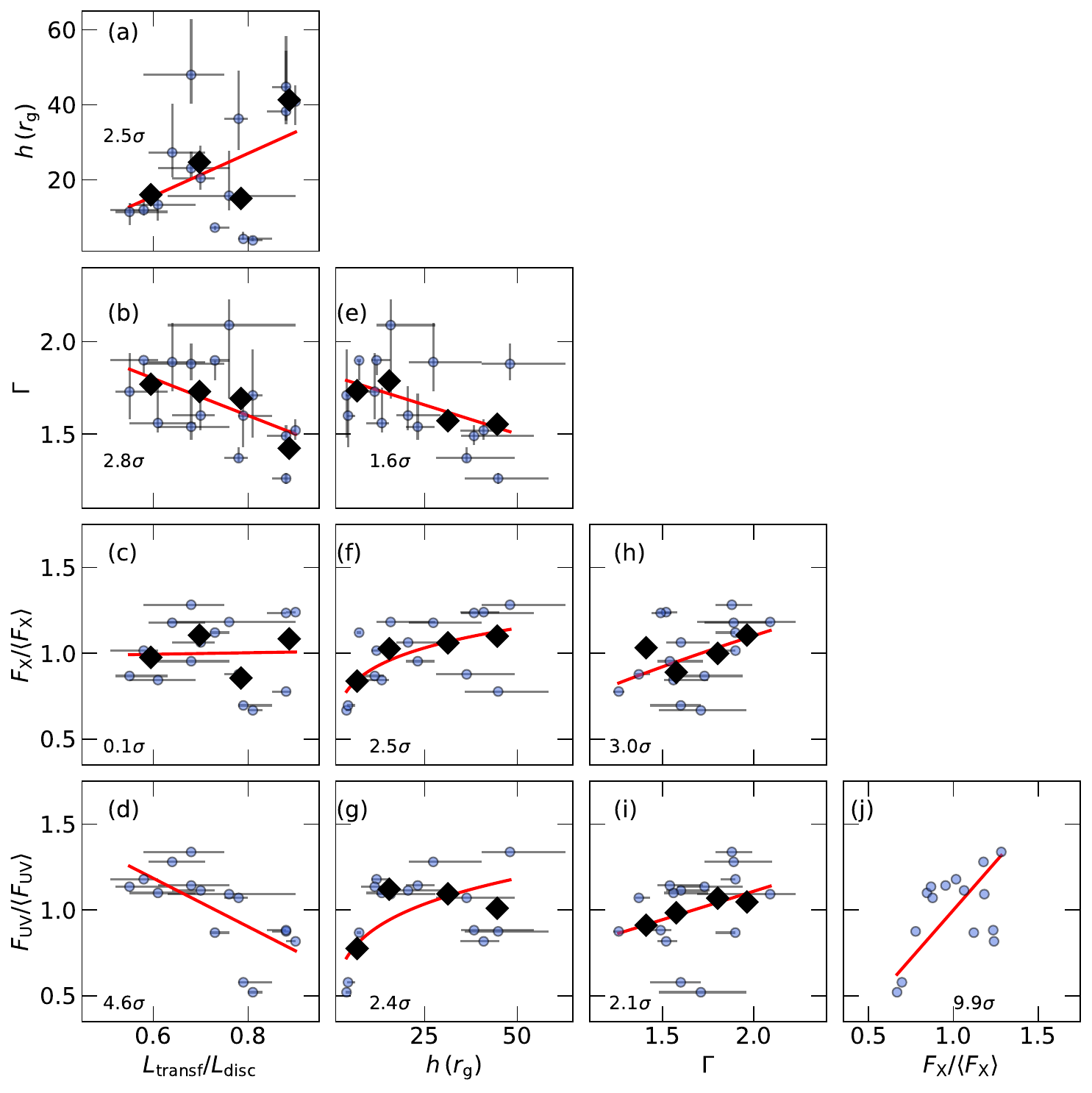}
    \caption{Correlations between various best-fit parameters, as determined by fitting the variable SEDs. Solid red lines show the best-fit straight line or power law model. {Black diamonds show the binned data.} Numbers in each plot indicate the significance of each slope not being zero (see Sect.\,\ref{sec:correlation}).}
    \label{fig:triangle_plot}
\end{figure}

We show in Fig.\,\ref{fig:triangle_plot} correlations between various best-fit corona parameters as well as the model X-ray and UV fluxes. Such correlations, if real, can help us investigate the origin of the observed variability. In other terms, these correlations can help us identify the primary physical parameters that are responsible for the observed variability in the X-rays and in the UV/optical bands. 

In order to search for significant correlations between the various parameters that are plotted in Fig.\,\ref{fig:triangle_plot}, we fitted a line (either in the linear or in the logarithmic space) and we then check if the best-fit slope is consistent with zero. If the slope is significantly different than zero, this would imply that there is a significant correlation between the two parameters. We used the ordinary least-squares (OLS) regression method \citep{Isobe90} of Y against X, OLS(Y|X), for all the relations except for the relation between the X-ray and UV flux (Panel j). Since neither the X-ray nor the UV flux can be consider as the ``effect'' and/or the ``cause'' of the other, it is better to treat these variables symmetrically. For that reason, we follow \cite{Isobe90} and we use the bisector when fitting the respective plot in Fig.\,\ref{fig:triangle_plot}. The best-fit lines are shown as red solid lines in Fig.\,\ref{fig:triangle_plot}, and the best-fit slopes are listed in Table\,\ref{tab:linefit}. 

We also report the significance of the best-fit slopes being non-zero in each of the panels, $\sigma_{\rm slope}$, which should be indicative of the significance of the correlation between the various parameters plotted in Fig.\,\ref{fig:triangle_plot}. Of course this is true only if the best-fit line does indeed provide a good fit to the data (the \cite{Isobe90} method already assumes that there is a linear relation between the two variables). However, the various plots in Figure \ref{fig:triangle_plot} show that a simple line (either in the linear or in the logarithmic space) may not provide a good fit to the data. The scatter between the data points and the best-fit lines in many plots is quite large. Significant outliers exist, which makes it difficult to even consider (let alone prove) whether there is a correlation or not. We suspect that most of the best-fit parameter values depend on more than one of the other physical parameters, and it is for this reason that the correlation between most parameters appears to be rather messy. 

For these reasons, we computed the Spearman's rank correlation coefficient, $\rho_{\rm S}$,  for all the correlations plotted in the same figure (results are listed in the last column of Table\,\ref{tab:linefit}), and we also binned the data in the various panels in Fig.\,\ref{fig:triangle_plot} (except Panels (d) and (j) where $\sigma_{\rm slope}$ and/or the correlation coefficient imply a strong correlation between the parameters). The binned points should indicate the mean correlation between the various parameters in each plot, and will allow us to investigate if they are correlated, at least on average (i.e. when the scatter of the points is taken off, to some extend). The black diamonds in these panels indicate the resulting mean correlation in each plot. In many cases, the binned data are consistent with the best fit lines. We consider this as an indication of an intrinsic correlation between various model parameters. 

For example, Panels\,(b) and (e) show that the binned data are very close to the best-fit line. In Panel\,(b), both $\rho_{\rm S}$ and $\sigma_{\rm slope}$ indicate an intrinsic correlation between $\Gamma$ and \ltransf, while in Panel\,(e), $\sigma_{\rm slope}$ is small. This is most likely due to the fact that, even if there is an intrinsic linear relation between the two parameters, the slope is small and we cannot detect it with a high accuracy with the available data. The binned data are close to the best-fit lines in Panels\,(f) and (h) as well. In the first case, both $\sigma_{\rm slope}$ and $\rho_{\rm S}$ indicate that an intrinsic correlation may be real, while in the second case $\rho_{\rm S}$ is rather small. But this is not surprising, because the correlation coefficient measures the scatter of the points around the best-fit line to the data in the linear space, while the best-fit relation in this panel is non-linear. As for Panel\,(i), $\rho_{\rm S}$ implies a weak correlation, although the binned points are close to the best-fit line, most probably because of the presence of two points which are outliers in this plot. 

We discuss below some of the correlations in the cases when the binned data are close to the best-fit lines, and $\sigma_{\rm slope}$ and/or $\rho_{\rm S}$ suggest either significant (at the 3$\sigma$ level) or quite probable correlations between the parameters. These correlations indicate that the main driver of the X-ray variability could be \ltransf\ and the corona height, and can explain why the correlation between the observed X--ray and UV fluxes may not always be perfect. In fact if the X-ray source is powered by the accretion process, it is even possible that UV and X-rays will be anti-correlated.


Panel\,(d) in Fig.\,\ref{fig:triangle_plot} shows that the UV flux strongly anti-correlates with \ltransf: as the power given to the corona increases, the available disc luminosity (hence UV flux as well) decreases. At the same time, as \ltransf\ increases, the X--ray flux should also increase. This can lead to an anti-correlation between UV and X--rays, if \ltransf\ were the only or the major parameter to vary. This is what most likely happens between segments s8-s12 (i.e. between MJD 56760 and 56800, highlighted as red shaded area in Fig.\,\ref{fig:lightcurve}). Fig.\,\ref{fig:param_lc} shows that \ltransf\, is the main parameter that varies during this period. \ltransf\ is overall decreasing during this period, which explains why the UV flux is increasing, while the X-ray flux is decreasing.

{The spectrum photon index appears to negatively correlate with both \ltransf\ and $h$ (Panels\,(b) and (e) in Fig.\,\ref{fig:triangle_plot}). The first correlation could be due to the fact that part of the power that is transferred to the corona should be used to heat it up, hence large power should imply a flatter spectrum (as suggested by Panel\,(b)). At the same time, Panel\,(e) implies weak, but suggestive, correlation between $\Gamma$ and $h$. The spectrum becomes flatter (i.e. it is most likely that the corona becomes hotter) as the height increases. In this case, the solid angle of the corona subtended to the disc decreases (note that, as we commented in the previous section, the corona radius appears to be constant, irrespective of the height). Consequently, the number of photons that enter in the corona should also decrease, hence the corona cooling should decrease, and the temperature increases (causing the $\Gamma$ flattening).} 

{Interestingly, Panel\,(c) in Fig.\,\ref{fig:triangle_plot} shows that the observed X-ray flux, in the $2-7$\,keV band, is not correlated with \ltransf. This can be explained by the $\Gamma$-\ltransf\ anti-correlation in Panel\,(b): as the power given to the corona increases (from \ltransf$\sim 0.5$ to 0.9), $\Gamma$ flattens from $\sim 2.1$ to $\sim 1.25$. This $\Gamma$ variability will cause a significant variation in the observed X-ray flux in the $2-7\,\rm keV$ band, as Panel\, (h) shows (this is the `steeper when brighter' X-ray spectral variability pattern commonly observed in Seyfert galaxies). Therefore, as the power given to the corona increases, the spectral slope flattens, hence the observed X-ray flux in the $2-7$\,keV band decreases, and this results in an apparent lack of correlation between \ltransf\ and $F_{\rm X}$.} 

Our results are in agreement with the findings of \cite{Panagiotou2022ccf}. The authors demonstrated that the low correlation between the X-rays and the UV/optical in NGC\,5548, obtained by using the cross-correlation function (CCF), could be caused by the lack of a static source configuration, as is for example the case of a dynamic corona, accompanied by changes in the photon index. The authors concluded that the moderate X-ray/UV CCF does not contradict a scenario in which the UV/optical variability is driven by X-ray illumination of the accretion disc. In the current work, we show that not only that changes in height and photon index are required to reproduce the broadband variability but also a change in the heating power given to the corona is needed.

We note that, although our results fully support the hypothesis of X-ray illumination of the disc, we do not have a physical explanation for various aspects of the model. For example,  our results support the idea that \ltransf\ is the main driver of the X-ray variability. This implies that the radius below which the power is transferred to the corona varies. However, the mechanism that could produce these variations remains unknown. The processes that could transfer the power from the disc to the corona remain unknown as well. It could be probable that the corona is consistent with a failed jet as suggested by \cite{Ghisellini2004}. The corona could also be formed by magnetic reconnection which is also how the power could be transferred to the corona from the accretion disc \citep[see e.g.][]{DiMatteo1998,  Merloni2001res, Merloni2001}. {Additional models invoking magnetic field, such as the jet emitting disc (JED) standard accretion disc \citep[JED-SAD;][]{Ferreira2006, Petrucci2008}, can potentially explain the heating of the X-ray corona and its connection to magnetic field. However, to the best of our knowledge, this model has not been applied to the correlated X-ray/UV variability that is observed in AGN.}

\subsection{Implications for the time lags}
\label{sec:lags}
\begin{figure}
    \centering
    \includegraphics[width = 1\linewidth]{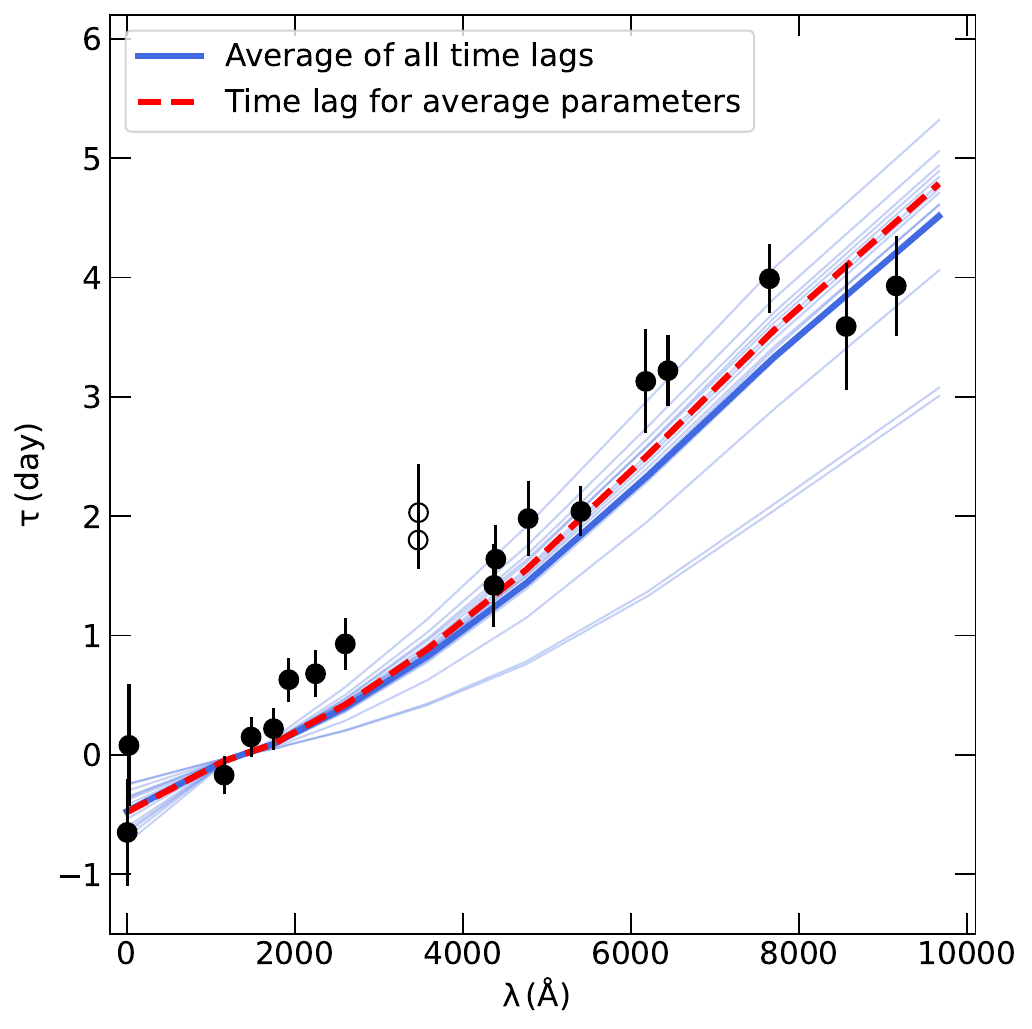}
    \caption{Model time lags obtained by considering the best-fit parameters from the 15 SEDs of NGC\,5548 (thin blue lines). The thick blue line correspond to the average of the 15 time-lag spectra. The dashed red line corresponds to the time lag spectrum obtained by considering the average values of the best-fitted parameters obtained by modelling the variable SEDs. The black circles are the observed continuum time lags of NGC\,5548. We show as empty black circles the measurements that are expected to be affected by the continuum emission from the broad-line region resulting in an excess in the delay.}
    \label{fig:lag}
\end{figure}

The observed X-ray and UV/optical flux variations in NGC\,5548 are due to continuous changes of the power that heats the corona (\ltransf) as well as its height. These two parameters strongly affect the UV/optical continuum responses of the accretion disc expected from thermal reverberation as demonstrated in \cite{Kammoun19lags,Kammoun21a} and K23. Therefore, we expect that the cross-correlation between the X-rays and the UV/optical bands, as well as the cross-correlation between the various UV/optical bands to continuously change in response to these variations in \ltransf\ and $h$. This implies that the time-lags between the various wavebands should also change.

In Fig.\,\ref{fig:lag} we show the observed time lags during the full STORM campaign as a function of wavelength \citep{Fausnaugh16}, hereafter the time-lag spectrum. We also show the model time-lag spectra expected during each of the 15 segments we analyse in this work, as thin blue lines. These time-lag spectra are estimated using the code of K23. We use as input parameters the values of $h, \Gamma$, and \ltransf\ listed in Table\,\ref{tab:results}. The average of the 15 time-lag spectra is shown as thick blue solid line. The red dashed line corresponds to the model time-lag spectrum when we assume the mean of the parameters listed in Table\,\ref{tab:results}, i.e. $h = 23.2\,\rg$, $\Gamma = 1.67$, and $\ltransf = 0.73$. The difference between the two lines is less than 10\% on average. This indicates that the time lags usually computed using long light curves are equivalent to the time lags of the mean of the parameters that determine the disc/corona system (i.e. $h$, \ltransf, \mdot, $\Gamma$). Our results are affected by the fact that we choose only a sub-sample of 15 SEDs. Thus the changes in the model parameters and consequently in the time lags is restricted to this sub-sample while in reality this may not be following the continuous changes in the source. Despite this caveat, the average time-lag spectrum and the time-lag spectrum obtained from the average parameters are both in good agreement with the observed time lags within $\sim 1.3\,\sigma$, $1.2\,\sigma$ on average, respectively. It is also worth noting that despite the fact that the variable SEDs modelled in this work are limited to wavelengths below $\sim 2000$\,\AA, the predicted time-lags describe well the observations up to much longer wavelength.

An important implication of these results is that, in reality, time lags are not constant. Thus, the measured time lags should depend on the duration of the light curves. As the length of observed light curves increases, the mean of the system parameters changes and so do the time lags. In this case, the observed time lags using various light curve segments should not be the same, but should indeed change with the light curve duration.

We have shown that the spectral and timing properties of NGC\,5548 during the STORM campaign could be well explained within the context of X-ray illumination of the accretion disc. Using this model we are able to explain the UV/optical continuum time-lag spectrum of this source, its power spectral density, as well as the time-averaged and the time-dependent SEDs, which are crucial tests for any model proposed to explain variability in a given source. We plan to repeat this kind of modelling of all the various available and suitable data sets of other sources in the future.


\section{Acknowledgements}

\begin{acknowledgements}
We would like to thank the anonymous referee for their comments which helped improve the clarity of the manuscript. EK acknowledges financial support from the Centre National d’Etudes Spatiales (CNES). EK and IEP acknowledge support from the European Union’s Horizon 2020 Programme under the AHEAD2020 project (grant agreement n. 871158). MD acknowledges support from GACR project 21-06825X and thanks for the institutional support from RVO:67985815.  This work made use of data supplied by the UK \textit{Swift} Science Data Centre at the University of Leicester. This research is based on observations made with the NASA/ESA \textit{Hubble Space Telescope} obtained from the Space Telescope Science Institute, which is operated by the Association of Universities for Research in Astronomy, Inc., under NASA contract NAS 5–26555. This work makes use of Matplotlib \citep{Matplotlib}, NumPy \citep{Numpy}, and SciPy \citep{Scipy}.

\end{acknowledgements}



\bibliographystyle{aa} 
\bibliography{references} 

\begin{appendix} 
\section{Best-fit results to the time-resolved SEDs}

\begin{table*}
    
    \caption{Best-fit results obtained by fitting the individual spectra as presented in Sect.\,\ref{sec:fitting}. We also show in the bottom line the average of the best-fitted parameters.}
    \label{tab:results}
    \begin{threeparttable}
    
    \begin{tabular}{llllllllll}
    \hline \hline \\ [-0.15cm]
	&	Time 	&	\ltransf & $h$			&	$\Gamma$										&	$N_{\rm H}$					&	$\log  \xi$ &	$R_{\rm C}$&$\rm \chi^2/dof$ &	$p_{\rm null}$	\\[0.15cm]
	&	(MJD)	&	 & (\rg)			&											&	$(10^{22}\,\rm cm^{-2})$					&	$ (\rm erg~cm~s^{-1})$ &(\rg)	 &	&	\\ \hline \\[-0.1 cm]
s1 & $56712.9 - 56714.2$ & $0.90_{-0.02}^{u}$ & $40.9_{-6.3}^{+4.3}$ & $1.52_{-0.05}^{+0.06}$ & $2.56_{-0.50}^{+0.93}$ & $1.31_{-0.17}^{+0.21}$ & 15.4 & 15.0/20 & 0.77 \\ [0.15cm]
s2 & $56721.8 - 56723.2$ & $0.81_{-0.01}^{+0.02}$ & $3.9_{-0.3}^{+1.2}$ & $1.71_{-0.23}^{+0.25}$ & $3.79_{-1.74}^{+1.32}$ & $1.29_{-0.13}^{+0.19}$ & 7.3 & 20.0/13 & 0.09 \\ [0.15cm]
s3 & $56731.6 - 56732.9$ & $0.79_{-0.01}^{+0.06}$ & $4.3_{-0.3}^{+2.0}$ & $1.60_{-0.17}^{+0.11}$ & $1.64_{-0.46}^{+2.51}$ & $0.59_{-3.59}^{+0.98}$ & 5.3 & 29.0/15 & 0.02 \\ [0.15cm]
s4 & $56740.0 - 56741.3$ & $0.76_{-0.13}^{+0.14}$ & $15.8_{-3.9}^{+12.0}$ & $2.09_{-0.40}^{+0.14}$ & $2.04_{-0.21}^{+0.54}$ & $0.83_{-0.26}^{+0.70}$ & 27.0 & 14.5/13 & 0.33 \\ [0.15cm]
s5 & $56747.9 - 56749.2$ & $0.64_{-0.05}^{+0.07}$ & $27.3_{-6.6}^{+13.1}$ & $1.89_{-0.16}^{+0.21}$ & $2.47_{-0.60}^{+0.69}$ & $1.39_{-0.22}^{+0.27}$ & 27.0 & 31.0/32 & 0.50 \\ [0.15cm]
s6 & $56752.7 - 56754.0$ & $0.88_{-0.03}^{+0.01}$ & $44.8_{-9.0}^{+13.6}$ & $1.26_{-0.03}^{+0.03}$ & $1.00_{-0.34}^{+0.32}$ & $1.32_{-0.35}^{+0.21}$ & 7.4 & 40.0/33 & 0.19 \\ [0.15cm]
s7 & $56755.7 - 56757.0$ & $0.88_{-0.04}^{u}$ & $38.3_{-3.4}^{+16.2}$ & $1.49_{-0.05}^{+0.06}$ & $1.31_{-0.29}^{+1.13}$ & $1.34_{-0.16}^{+0.38}$ & 13.1 & 53.7/36 & 0.03 \\ [0.15cm]
s8 & $56758.8 - 56760.0$ & $0.73_{-0.01}^{+0.03}$ & $7.3_{-0.6}^{+0.5}$ & $1.90_{-0.11}^{u}$ & $1.88_{-0.16}^{+0.45}$ & $1.09_{-0.24}^{+0.16}$ & 11.5 & 40.9/33 & 0.16 \\ [0.15cm]
s9 & $56769.7 - 56771.0$ & $0.58_{-0.07}^{+0.03}$ & $12.0_{-1.4}^{+3.8}$ & $1.90_{-0.08}^{u}$ & $2.99_{-0.63}^{+0.32}$ & $1.74_{-0.18}^{+0.18}$ & 14.7 & 35.0/22 & 0.04 \\ [0.15cm]
s10 & $56779.2 - 56780.0$ & $0.70_{-0.06}^{+0.03}$ & $20.5_{-3.1}^{+8.7}$ & $1.60_{-0.08}^{+0.16}$ & $2.85_{-0.99}^{+0.70}$ & $1.64_{-0.32}^{+0.22}$ & 10.3 & 27.9/33 & 0.72 \\ [0.15cm]
s11 & $56788.0 - 56789.0$ & $0.61_{-0.06}^{+0.08}$ & $13.4_{-4.2}^{+1.9}$ & $1.56_{-0.05}^{+0.19}$ & $2.18_{-0.84}^{+1.08}$ & $1.48_{-0.29}^{+0.34}$ & 6.6 & 47.9/29 & 0.02 \\ [0.15cm]
s12 & $56798.7 - 56800.0$ & $0.55_{-0.03}^{+0.08}$ & $11.5_{-3.6}^{+2.4}$ & $1.73_{-0.15}^{+0.21}$ & $2.93_{-0.67}^{+0.49}$ & $1.29_{-0.15}^{+0.14}$ & 9.4 & 30.1/29 & 0.40 \\ [0.15cm]
s13 & $56810.7 - 56812.0$ & $0.68_{-0.07}^{+0.08}$ & $23.1_{-3.2}^{+4.5}$ & $1.54_{-0.07}^{+0.18}$ & $1.81_{-0.42}^{+1.60}$ & $1.09_{-0.33}^{+0.42}$ & 9.3 & 27.9/21 & 0.14 \\ [0.15cm]
s14 & $56823.4 - 56824.7$ & $0.68_{-0.10}^{+0.07}$ & $48.0_{-7.7}^{+14.9}$ & $1.88_{-0.09}^{+0.11}$ & $2.41_{-1.14}^{+1.06}$ & $1.26_{-0.56}^{+0.24}$ & 41.5 & 10.6/13 & 0.65 \\ [0.15cm]
s15 & $56827.2 - 56828.5$ & $0.78_{-0.03}^{+0.02}$ & $36.3_{-8.3}^{+12.9}$ & $1.37_{-0.02}^{+0.06}$ & $1.15_{-0.17}^{+0.33}$ & $1.09_{-0.19}^{+0.20}$ & 8.6 & 41.3/34 & 0.18 \\ [0.15cm]

\hline \\[-0.15cm]
Mean & & 0.73 & 23.2 & 1.67 & 2.20 & 1.25 & 14.3 & & \\
\hline
    \end{tabular}
    \begin{tablenotes}
        \item[$^u$] Upper limit.
    \end{tablenotes}
    
    \end{threeparttable}
\end{table*}

\end{appendix}

\end{document}